\documentclass[prl,reprint,amsmath,amssymb,aps,floatfix]{revtex4-2}
\usepackage[margin=0.75in]{geometry}
\usepackage[utf8]{inputenc}
\usepackage{xr}
\usepackage{xr-hyper}
\usepackage{amsmath}
\usepackage{amssymb,float}

\usepackage{soul}
\usepackage[normalem]{ulem}
\usepackage{amsfonts}
\usepackage{amstext}
\usepackage{appendix}
\usepackage{mathtools}
\usepackage{amsthm}
\usepackage{subcaption}
\usepackage{graphicx}
\usepackage{color}
\usepackage{lipsum}
\usepackage{natbib}
\usepackage{xcolor}
\usepackage{float}
\usepackage{bm}
\usepackage[hidelinks]{hyperref}
\usepackage{comment}
\usepackage[normalem]{ulem}
\usepackage{ragged2e}
\usepackage[font={scriptsize}, skip=0pt,format=plain,singlelinecheck=false]{caption}

\newcommand{\wo}{\setminus}

\usepackage[nottoc,notbib]{tocbibind}
\usepackage{helvet}
\def\be{\begin{equation}}
\def\ee{\end{equation}}
\def\bea{\begin{eqnarray}}
\def\eea{\end{eqnarray}}

\def\be{\begin{equation}}
\def\ee{\end{equation}}
\def\bea{\begin{eqnarray}}
\def\eea{\end{eqnarray}}

\def\a{\alpha}

\begin{document}
\title{\hspace*{-0.45cm}Energy flow controls the stability of multitrophic ecosystems with stratified nonreciprocity\hspace*{-0.5cm}}

\author{Rukmani Ramachandran}
\author{Akshit Goyal} 
\email{akshitg@icts.res.in}
\address{International Centre for Theoretical Sciences, Tata Institute of Fundamental Research, Bengaluru 560089.}

\begin{abstract} \noindent Complex systems with nonreciprocal interactions are often stratified into layers. Ecosystems are a prime example, where species at one trophic level grow by consuming those at another. Yet the dynamical consequences of such stratified nonreciprocity—where the correlation between growth and consumption differs across trophic levels—remain unexplored.
Here, using an ecological model with three trophic levels, we reveal an emergent asymmetry: nonreciprocal interactions between consumers and predators (top and middle level) destabilize ecosystems far more readily than nonreciprocity between consumers and resources (middle and bottom level).
We analytically derive the phase diagram for the model and show that its stability boundary is controlled by energy flow across trophic levels. 
Because energy flows upward---from resources to predators---diversity is progressively lower at higher trophic levels, which we show explains the asymmetry.
Lowering energy flow efficiency flips the asymmetry toward resources and remarkably expands the stable region of the phase diagram, suggesting that the famous ``10\% energy transfer'' seen in natural ecosystems might promote stability. 
More broadly, our findings show that the location of nonreciprocity within a complex network, not merely its magnitude, determines stability.
\end{abstract}

\maketitle

\noindent Nonreciprocal interactions---where $A$ affects $B$ differently than $B$ affects $A$---are ubiquitous in complex systems and can drastically impact their dynamics~\cite{you2020nonreciprocity,fruchart2021non,dinelli2023non,belyansky2025phase,kryuchkov2018dissipative,tan2022odd}. Most theoretical works assume a fixed degree of reciprocity common to all interacting pairs in a system (though see \cite{weis2025generalized}). Yet the degree of reciprocity need not be fixed and may vary across a system. This is especially true for a variety of systems which are structured into layers: from neural networks with feedforward and feedback pathways~\cite{angelucci2017circuits,pennartz2019towards,markov2014weighted} to supply chains with asymmetric upstream and downstream dependencies~\cite{pu2023dependence,gelderman2003handling}.

Ecosystems serve as a prime example of such complex systems. They are organized into trophic levels of resources, consumers and predators~\cite{odum1968energy,lindeman1942trophic,rigler1975concept,hairston1993cause}, where species in one trophic level interact with those in adjacent trophic levels. The resulting ecological interaction networks can also be nonreciprocal, where the degree of reciprocity is often measured by the correlation between growth and consumption rates at each interface between two trophic levels~\cite{blumenthal2024phase,rowland2025resource}. The correlation between how much consumers grow and how much they deplete resources (middle and bottom level) may differ substantially from the corresponding correlation between consumer and predators (top and middle level). This suggests that in ecosystems, reciprocity can differ---or be stratified---across trophic levels.


While such ``stratified nonreciprocity'' is likely a generic feature of ecosystems, we do not understand its consequences for their dynamics. Recent statistical physics approaches have made significant progress understanding ecological dynamics in systems with many species~\cite{BuninGLV, advani2018statistical, cui2020effect, mahadevan2023spatiotemporal, pearce2020stabilization, patro2025emergent,arnoulx2024many,mallmin2024chaotic}. These works have established that below a critical reciprocity, dynamics become unstable and transition to chaos~\cite{blumenthal2024phase,liu2025complex,rowland2025resource,hu2022emergent}. However, they have largely analyzed ecosystems with only one or two trophic levels (though see \cite{feng2024emergent,liu2024ecosystem}), where there is only one measure of reciprocity. With three or more trophic levels, there can be multiple measures of reciprocity stratified across trophic levels. Yet we do not know whether nonreciprocal interactions at the top trophic level control stability the same way as at the bottom.

In this Letter, we analyze the stability of a nonreciprocal multitrophic ecological model using analytic theory and simulations. We find an emergent asymmetry: ecosystems can be destabilized much more easily by nonreciprocal interactions with predators (top) than resources (bottom). As a result, dynamics under nonreciprocal consumer-resource interactions, which would otherwise be unstable, can be stabilized if consumer-predator interactions are reciprocal---but \textit{not} vice versa. 

We show that the efficiency of energy flow across trophic levels governs this asymmetry. Decreasing efficiency transitions ecosystems to the opposite regime, where they instead become more easy to destabilize through resources. 
Inefficient energy flow also vastly expands the stable region, suggesting that the ``10\% law'' in ecology------that $\sim$10\% of energy is transferred between successive trophic levels~\cite{lindeman1942trophic,odum1968energy}---helps to stabilize ecosystems. 
Our results reveal how energy flow shapes the dynamics of multilayered ecosystems with nonreciprocal interactions.

\begin{figure}[ht!]    
   \centering\includegraphics[width=0.9\columnwidth]{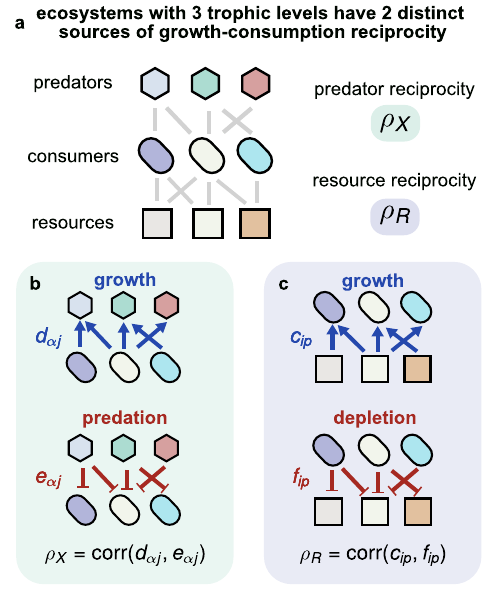}
    \caption{\justifying\textsf{\textbf{Stratified nonreciprocity in a multitrophic ecosystem.}
(a) Schematic of an ecosystem with three trophic levels: resources (squares), consumers (ellipses), and predators (hexagons). The ecosystem has two distinct reciprocity parameters: predator reciprocity $\rho_X$ and resource reciprocity $\rho_R$.
(b) At the consumer-predator interface, growth benefits $d_{\alpha j}$ (blue) and predation impacts $e_{\alpha j}$ (red) are correlated with strength $\rho_X = \text{corr}(d, e)$.
(c) At the resource interface, growth benefits $c_{ip}$ (blue) and resource depletion $f_{ip}$ (red) are correlated with strength $\rho_R = \text{corr}(c, f)$. 
}}
\label{fig:fig1}
\end{figure}


\textit{Setup.---}We consider an ecosystem with three trophic levels: $M_R$ resource species, $M_N$ consumer species, and $M_X$ predator species (Fig.~\ref{fig:fig1}a). The dynamics are governed by the following set of equations:
\begin{align}
\frac{dX_\alpha}{dt} &= X_\alpha \left( \eta \sum_{j=1}^{M_N} d_{\alpha j} N_j - u_\alpha \right), \label{eq:dyn_X} \\[0.8em]
\frac{dN_i}{dt} &= N_i \left( \eta \sum_{p=1}^{M_R} c_{ip} R_p - m_i - \sum_{\beta=1}^{M_X} e_{\beta i} X_\beta \right), \label{eq:dyn_N} \\[0.8em]
\frac{dR_p}{dt} &= R_p \left( \kappa_p - R_p - \sum_{j=1}^{M_N} f_{jp} N_j \right), \label{eq:dyn_R}
\end{align}
where $X_\alpha$, $N_i$, and $R_p$ denote the abundances of predator $\alpha$, consumer $i$, and resource $p$, respectively. The interaction matrices encode both species growth and resource consumption at each interface between trophic levels. At the resource interface, $c_{ip}$ is the energy flux that consumer $i$ gains from resource $p$, while $f_{ip}$ is how much consumer $i$ depletes resource $p$. At the predator interface, $d_{\alpha j}$ is the energy flux that predator $\alpha$ gains from consumer $j$, while $e_{\alpha j}$ is how much predator $\alpha$ depletes consumer $j$. The parameters $u_\alpha$ and $m_i$ represent the mortality rates of predator $\a$ and consumer $i$ respectively, and $\kappa_p$, the supply rate of resource $\a$. The parameter $\eta \leq 1$ is the energy flow efficiency, representing the fraction of energy consumed at a lower trophic level that is converted to growth at a higher trophic level~\cite{welch1968relationships,pauly1995primary}.

\textit{Stratified nonreciprocity.---}With only two trophic levels---consumers and resources---a single parameter $\rho$ captures nonreciprocity: the correlation between consumer growth and resource consumption~\cite{blumenthal2024phase,rowland2025resource,liu2025complex}. In an ecosystem with three trophic levels, there are two distinct reciprocity parameters: $\rho_R$ at the resource level and $\rho_X$ at the predator level (Fig.~\ref{fig:fig1}b--c). These ecosystems are thus the simplest ones that exhibit stratified nonreciprocity, where the correlation between species growth and consumption can differ across trophic levels. 


While na\"ively one might guess that both reciprocity parameters will have similar effects, we will show that this is not the case. Directional energy flow breaks the symmetry by progressively reducing diversity up the food web. 
This gradient in diversity across trophic levels ultimately determines whether predator reciprocity $\rho_X$ or resource reciprocity $\rho_R$ controls stability.



\textit{Thermodynamic limit and cavity method.---}We work in the limit of highly diverse species pools, with $M_R, M_N, M_X \to \infty$ such that their ratios, $r_1 = M_N/M_R$ and $r_2 = M_X/M_N$, are finite. 
Following the statistical physics tradition in ecology~\cite{MayStability,BuninGLV, advani2018statistical, cui2020effect,mahadevan2023spatiotemporal,pearce2020stabilization}, we treat the large number of microscopic parameters as quenched random variables, allowing us to characterize the typical behavior of ecosystems in this high-diversity limit.
Carrying capacities and mortality rates are drawn as $\kappa_p = \kappa + \sigma_\kappa \delta\kappa_p$, $m_i = m + \sigma_m \delta m_i$, and $u_\alpha = u + \sigma_u \delta u_\alpha$, where $\delta\kappa_p$, $\delta m_i$, and $\delta u_\alpha$ are independent standard normal variables. The interaction matrices are parameterized to ensure a proper thermodynamic limit as follows:
\begin{align}
c_{ip} &= \frac{\mu_c}{M_R} + \frac{\sigma_c}{\sqrt{M_R}} \xi_{ip}, \\[0.5em]
f_{ip} &= \frac{\mu_f}{M_R} + \frac{\sigma_f}{\sqrt{M_R}} \left( \rho_R \xi_{ip} + \sqrt{1 - \rho_R^2} \, 
\zeta_{ip} \right),
\end{align}
where $\xi_{ip}$ and $\zeta_{ip}$ are uncorrelated standard normal variables. We repeat this analogously for the consumer-predator layer, with interaction parameters $d_{\a j}$ and $e_{\a j}$ with means $\mu_d/M_N$ and $\mu_e/M_N$, standard deviations $\sigma_d/\sqrt{M_N}$ and $\sigma_e/\sqrt{M_N}$ respectively, and correlation $\rho_X$.
The correlation parameters $\rho_R = \text{corr}(c_{ip}, f_{ip})$ and $\rho_X = \text{corr}(d_{\alpha j}, e_{\alpha j})$ thus independently control the degree of reciprocity at each trophic interface (Fig.~\ref{fig:fig1}). Throughout this work, simulations use $M_R = M_N = M_X = 100$ species per layer, with $\mu_c = \mu_f = \mu_d = \mu_e = 100$, $\sigma_c = \sigma_f = \sigma_d = \sigma_e = 4$, and $\kappa = m = u = 1$ with $\sigma_\kappa = \sigma_m = \sigma_u = 0.1$. This choice of parameters ensures that all trophic levels are statistically equivalent, allowing us to focus only on nonreciprocity and energy flow. Dynamics are integrated using an adaptive stiff-solver scheme~\cite{petzold1983automatic,hindmarsh1983odepack} with a small immigration rate to regularize extinctions (Appendix~A).

\begin{figure*}[ht!]    
   \centering\includegraphics[width=\textwidth]{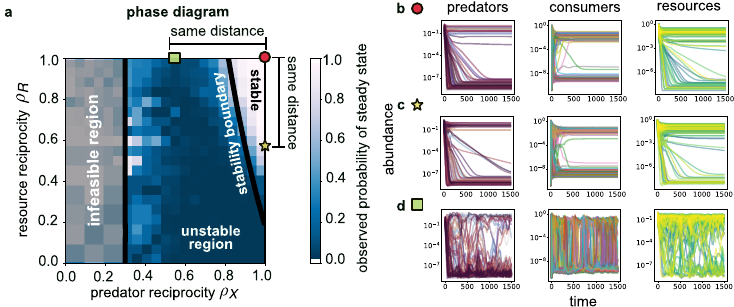}
    \vspace{1pt}\caption{\justifying\textsf{\textbf{Ecosystem stability relies more on reciprocal interactions with predators than resources. 
    }
(a) Phase diagram in the $(\rho_X, \rho_R)$ plane showing stable (white), unstable (purple), and infeasible (light blue) regions. The analytically-derived stability boundary (black curve) is asymmetric: reducing $\rho_X$ leads to instability earlier than reducing $\rho_R$. Green square and yellow star mark points equidistant from the reciprocal corner. Heatmap shows probability of reaching steady state from numerical simulations. Gray indicates infeasible region where the cavity solutions admit no nontrivial solution; the infeasibility boundary, obtained analytically, is next to it in black (Appendix~D).
(b) Dynamics at the reciprocal point ($\rho_X = 1$, $\rho_R = 1$; circle): all three trophic layers (predators, consumers, producers) reach stable steady states.
(c) Dynamics with reduced resource reciprocity ($\rho_X = 1$, $\rho_R = 0.5$, star): the ecosystem remains stable despite nonreciprocal consumer-resource interactions.
(d) Dynamics with reduced predator reciprocity ($\rho_X = 0.5$, $\rho_R = 1$, square): the ecosystem is unstable, exhibiting chaotic fluctuations across all trophic levels. The asymmetry between (c) and (d) demonstrates inherent top-down control of stability.}}
\label{fig:fig2}
\end{figure*} 

To analyze the thermodynamic limit analytically, we employ the cavity method from the statistical physics of disordered systems~\cite{advani2018statistical, cui2020effect, BuninGLV,blumenthal2024phase,feng2024emergent}. The method proceeds by considering how ecosystem properties change upon the addition of a single new species to each trophic level. In the limit of large species pools, this perturbation is small and can be treated self-consistently, yielding closed-form equations for observables: the fraction of species surviving at each trophic level $\phi_R$, $\phi_N$, and $\phi_X$, the mean and fluctuations of species abundances $\langle R\rangle$, $\langle N\rangle$, $\langle X\rangle$, $\langle R^2\rangle$, $\langle N^2\rangle$, $\langle X^2\rangle$, and the susceptibilities $\nu^{(N)}$, $\chi^{(X)}$ and $K^{(R)}$ that characterize how abundances respond to changes in environmental parameters (for the full calculation, see Appendices~B--C). 

\textit{Stability depends on predators more than resources.---}With full reciprocity $\rho_R=\rho_X = 1$, dynamics follow an optimization principle and are guaranteed to reach a stable fixed point~\cite{marsland2020minimum,mehta2019constrained,feng2024emergent}. When there is nonreciprocity, stability is not guaranteed and there is instead a critical boundary $(\rho_X^\star,\rho_R^\star)$ that separates stable from unstable ecosystems. We find that this boundary is asymmetric---more sensitive to predators than resources. To illustrate this, we consider two points in the $(\rho_X, \rho_R)$ plane equidistant from the fully reciprocal corner $\rho_X=1, \rho_R = 1$ (red circle in Fig.~\ref{fig:fig2}a).  At full reciprocity, ecosystems reach a stable steady state (Fig.~\ref{fig:fig2}b). However, when we move the same distance from this point in both directions, we find that the dynamics are quite different. When we reduce only resource reciprocity to $\rho_R = 0.5$, dynamics are still stable: species abundances relax to a fixed point (Fig.~\ref{fig:fig2}c). In contrast, reducing only predator reciprocity to $\rho_X = 0.5$ destabilizes the ecosystem, resulting in chaotic abundance fluctuations across all trophic levels (Fig.~\ref{fig:fig2}d). This is despite all trophic levels being statistically equivalent in terms of disorder parameter values. These examples suggest that at perfect energy flow $\eta=1$, ecosystems are more sensitive to nonreciprocity at the top trophic level.

Using the cavity method, we can analytically compute the phase boundary between stable and unstable phases and derive this asymmetry (Appendix~E). Following Refs.~\cite{BuninGLV, blumenthal2024phase}, we perturb the steady-state abundances of surviving species at each trophic level and compute response functions---random variables encoding species abundance changes at each layer to small perturbations. The loss of stability is signaled by the divergence of these responses' second moments, marking the breakdown of replica symmetry. 
This results in an instability condition that couples all three trophic layers through the determinant of a $3 \times 3$ matrix (Appendix~E). The stability boundary is a curve in the $(\rho_X, \rho_R)$ plane given by:
\begin{equation}
(\rho_R^\star)^2 \phi_R \left[(\rho_X^\star)^2 \phi_N - \phi_X\right] = (\rho_X^\star)^2 (\phi_X - \phi_N)^2,
\label{eq:instab_boundary_main}
\end{equation}
where $\phi_R = M^*_R/M_R$, $\phi_N = M^*_N/M_N$, and $\phi_X = M^*_X/M_X$ are the surviving fractions of resources, consumers, and predators, respectively. These $\phi$'s encode how much of the diversity in the species pool coexists at each trophic level. Eq.~\eqref{eq:instab_boundary_main} remarkably shows that stability is controlled entirely by species diversity patterns across trophic levels. Any other parameters, such as microscopic interaction strengths, can only affect stability by influencing these $\phi$'s. The analytical boundary in Eq.~\eqref{eq:instab_boundary_main} agrees well with numerical simulations, shown via the probability of reaching steady state (Fig.~\ref{fig:fig2}a).

\begin{figure}[ht!]    
   \centering\includegraphics[width=\columnwidth]{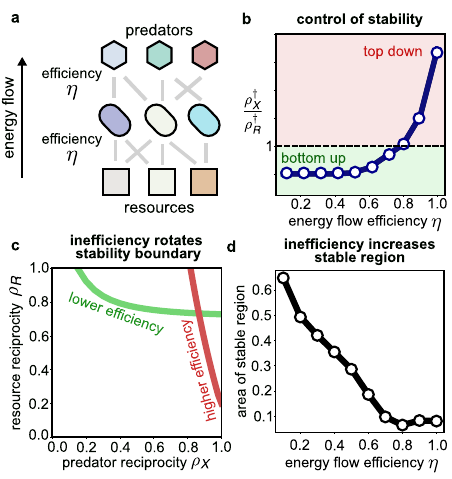}
    \vspace{0.05cm}\caption{\justifying\textsf{\textbf{Energy flow efficiency controls the stability boundary.}
(a) Schematic of energy flow across an ecosystem: a factor $\eta$ of the energy at one trophic level is transferred to the next higher level.
(b) Ratio of critical reciprocities $\rho_X^\dagger/\rho_R^\dagger$ as a function of $\eta$. When the ratio exceeds 1 (pink), stability is top-down controlled; when below 1 (green), bottom-up controlled.
(c) The ecosystem stability boundary rotates as energy flow efficiency $\eta$ decreases. At high efficiency ($\eta = 1$, red), the boundary is steep; at low efficiency ($\eta = 0.6$, green), the boundary is shallow. Shown are analytic solutions.
(d) Total area of the stable region increases as $\eta$ decreases, showing that energy dissipation is stabilizing.
}}
\label{fig:fig3}
\end{figure}

To characterize the asymmetry quantitatively, we compute the critical reciprocity at each interface when the other is fully reciprocal. When $\rho_R^\star = 1$, instability occurs at $\rho_X^\dagger$; when $\rho_X^\star=1$, instability occurs at $\rho_R^\dagger$. Using the cavity instability Eq.~\eqref{eq:instab_boundary_main}, we find:
\be
\rho_X^\dagger = \sqrt{\frac{\phi_R \phi_X}{\phi_R \phi_N - (\phi_X - \phi_N)^2}}, \quad \rho_R^\dagger = \sqrt{\frac{\phi_N - \phi_X}{\phi_R}}.
\label{eq:rhoX_rhoR_crit}
\ee
Once again, the critical reciprocities $\rho_X^\dagger$ and $\rho_R^\dagger$, which represent the corners of the instability boundary in Fig.~\ref{fig:fig2}a, are controlled entirely by the diversity $\phi$ at each level. Because energy flows upward through the food web, diversity generically decreases with trophic levels: $\phi_R \ge \phi_N \ge \phi_X$.
This intrinsic diversity gradient biases ecosystems toward being more sensitive to nonreciprocity at higher trophic levels. We quantify this sensitivity by the ratio:
\begin{equation}
\frac{\rho_X^\dagger}{\rho_R^\dagger} 
= 
\sqrt{
\frac{\phi_R^2 \phi_X}{
(\phi_N - \phi_X)\left[\phi_R \phi_N - (\phi_X - \phi_N)^2\right]
}}.
\label{eq:asymmetry}
\end{equation}
When this ratio $>1$, the ecosystem is more sensitive to nonreciprocity at the top (predator-consumer interface); when it is $<1$, the ecosystem is more sensitive to the bottom (consumer-resource interface). In ecology, these regimes are called top-down control and bottom-up control, respectively. While they typically refer to which trophic level regulates biomass, here we find that similar regimes also apply to stability.
In these terms, at perfect energy flow $\eta=1$, since $\rho_X^\dagger/\rho_R^\dagger > 1$, stability is controlled top-down (Fig.~\ref{fig:fig2}).

\textit{Energy flow controls stability boundary.---}While the direction of energy flow creates an asymmetric sensitivity to nonreciprocity, we find that the asymmetry is not fixed and can be modulated. 
In particular, the efficiency of energy flow $\eta$ across trophic levels turns out to be a key control parameter (Fig.~\ref{fig:fig3}a). The efficiency $\eta$ modulates the asymmetric sensitivity ratio $\rho_X^\dagger/\rho_R^\dagger$ implicitly through diversities $\phi_R$, $\phi_N$, and $\phi_X$. Increasing energy dissipation by reducing $\eta$ intensifies competitive exclusion at every trophic step. This steepens the diversity gradient $\phi_R > \phi_N > \phi_X$ already present when $\eta = 1$ (Appendix~F).

The steeper diversity gradient has a dramatic effect on the stability boundary. To see this most clearly, notice that at low efficiency $\eta\ll1$, the diversity gradient becomes so steep that predator diversity nearly collapses, i.e., $\phi_X\ll\phi_N$ while $\phi_R\sim\mathcal{O}(1)$. Here, Eq.~\eqref{eq:asymmetry} reduces to $\rho_X^\dagger/\rho_R^\dagger\approx\sqrt{\phi_R\phi_X}/{\phi_N}\ll 1$. Hence at low efficiency, the ecosystem becomes more sensitive to nonreciprocal interactions through resource than predators. In ecological terminology, stability transitions from top-down to bottom-up control.
Plotting $\rho_X^\dagger/\rho_R^\dagger$ as a function of $\eta$ using our cavity solutions confirms this transition (Fig.~\ref{fig:fig3}b). Geometrically, reducing $\eta$ causes the stability boundary to rotate in the $(\rho_X,\rho_R)$ plane (Fig.~\ref{fig:fig3}c) as the ratio $\rho_X^\dagger/\rho_R^\dagger$ decreases. 
By visualizing the rotated stability boundary at low efficiency, we can see that it is now much easier to destabilize the ecosystem through resources, not predators. 

In addition to rotating the boundary, decreasing $\eta$ also translates it and expands the total area of the stable region (Fig.~\ref{fig:fig3}d). The stable area increases as $\eta$ decreases, reaching its maximum at the lowest values of $\eta$. This reveals that energy dissipation across trophic levels is intrinsically stabilizing: ecosystems that transfer energy inefficiently are more stable against nonreciprocal interactions than highly efficient ones. 

\textit{Discussion.---}Here, we analyzed ecosystems with multiple trophic levels, where the correlation between growth and consumption—a measure of reciprocity in ecological interactions—can differ at each interface. We found an emergent asymmetry: stability is far more sensitive to nonreciprocity at the top of the food web than at the bottom. Concretely, reducing reciprocity between consumers and predators destabilizes ecosystems much more readily than the same reduction between consumers and resources. These results generalize to $N>3$ trophic levels. Even in this case, stability depends on survival fractions or diversities $\phi$ at each trophic level, which decrease as you go up trophic levels (Appendix~H). This diversity gradient ultimately leads to an asymmetry in how ecosystems respond to nonreciprocity at different trophic levels. Taken together, our high-diversity model unifies both dynamic and energetic paradigms of  multitrophic ecosystems~\cite{barbier2019pyramids}.

A central finding is that energy flow efficiency acts as a control parameter that rotates and translates the stability boundary. As efficiency decreases, ecosystems transition from being more sensitive to nonreciprocal interactions with predators to being more sensitive to nonreciprocal interactions with resources---from top-down to bottom-up control of stability.
Remarkably, lowering efficiency also expands the stable region, suggesting that energy dissipation across trophic levels is intrinsically stabilizing. 
This result provides a new perspective on the ecological ``10\% law''~\cite{odum1968energy,lindeman1942trophic,rigler1975concept,hairston1993cause,pauly1995primary,welch1968relationships}---the empirical observation that roughly 10\% of energy is transferred between successive trophic levels. 
We speculate that the energy efficiency observed in natural ecosystems may reflect a balance between two effects: the stabilizing benefit of inefficiency, and the constraint that low efficiency places on the total number of sustainable trophic levels. 
Thus, the 10\% value may be self-organized to permit food webs which are both complex and stable. 

In the future, it would be valuable to test these predictions experimentally in ecosystems where trophic structure can be controlled and population dynamics monitored, e.g., in microbial communities with controllable phage and resources~\cite{hu2022emergent,dal2021resource,pyenson2024diverse,goldford2018emergent}. 
Further, while we focused on energy flow efficiency, it would be interesting to understand how other ecosystem parameters affect stability. Preliminary results suggest that one such key parameter is the relative ratio of species pool sizes across trophic levels, $r_1/r_2$. In other words, pyramidal diversity distributions are much more likely to yield stable ecosystems (Appendix~G). It is also crucial to understand how eco-evolutionary dynamics would alter the unstructured, disordered interaction network we considered here~\cite{feng2025theory,li2025population}, and its consequences for stability.

Finally, our findings have implications beyond ecology. Many complex systems are hierarchically structured into layers: neural networks, supply chains and social networks~\cite{pu2023dependence,gelderman2003handling,angelucci2017circuits,pennartz2019towards,markov2014weighted}. In such systems, nonreciprocity may differ at different interfaces, e.g., in feedforward and feedback connections along different layers of a neural network~\cite{weis2025generalized}. Our results suggest that the dynamical consequences of such nonreciprocity can depend strongly on \textit{where} in the network it occurs, not just on its magnitude.

\emph{Acknowledgements.---}We thank P. Mehta, S. Shankar, and M. Dal Bello for valuable discussions. We are grateful to P. Mehta for comments on an earlier draft of this manuscript. A.G. acknowledges support from the DAE under project no. RTI4001, an Ashok and Gita Vaish Junior Researcher Award, as well as a Ramanujan Fellowship.

\bibliography{ref}

\clearpage
\begin{widetext}

\begin{widetext}

\appendix

\captionsetup{font=normalsize}
\renewcommand{\theequation}{S\arabic{equation}}
\renewcommand{\thefigure}{S\arabic{figure}}
\setcounter{figure}{0}
\setcounter{section}{0}
\renewcommand{\thesection}{\Alph{section}}

\makeatother
{\hskip 150pt\textbf{SUPPLEMENTARY INFORMATION}}
\tableofcontents

\section*{Appendix A: Model setup and definitions}
\label{sec:model}

\subsubsection{Dynamical equations}

The objective of these appendices is to derive the steady-state behavior of a three-trophic-level ecosystem using the cavity method from statistical physics. We consider an ecosystem comprising $M_R$ resource species at the bottom trophic level, $M_N$ consumer (herbivore) species at the middle level, and $M_X$ predator (carnivore) species at the top level. The dynamics are governed by a coupled set of ordinary differential equations describing how each population changes over time.

The predator dynamics follow:
\begin{align}
\frac{dX_\alpha}{dt} &= X_\alpha \left( \eta\sum_j d_{\alpha j} N_j - u_\alpha \right),
&\quad \alpha = 1 \dots M_X
\end{align}
where $X_\alpha$ denotes the abundance of predator $\alpha$, $d_{\alpha j}$ represents the benefit predator $\alpha$ derives from consuming consumer $j$, $\eta$ is an energy conversion efficiency, and $u_\alpha$ is the mortality rate of predator $\alpha$.

The consumer dynamics are governed by:
\begin{align}
\frac{dN_i}{dt} &= N_i \left(\eta \sum_\alpha c_{i\alpha} R_\alpha - m_i - \sum_\beta e_{\beta i} X_\beta \right),
&\quad i = 1 \dots M_N
\end{align}
where $N_i$ is the population of consumer $i$, $c_{i\alpha}$ encodes the consumption preference of consumer $i$ for resource $\alpha$, $m_i$ is the mortality rate, and $e_{\beta i}$ describes the predation pressure from predator $\beta$ on consumer $i$.

Finally, the resource dynamics obey:
\begin{align}
\frac{dR_p}{dt} &= R_p \left( \kappa_p - R_p - \sum_j f_{jp} N_j \right),
&\quad p = 1 \dots M_R
\end{align}
where $R_p$ is the abundance of resource $p$, $\kappa_p$ is its carrying capacity, and $f_{jp}$ quantifies the depletion rate of resource $p$ by consumer $j$. The term $-R_p$ represents self-limiting logistic growth. This is consistent with resources being biological, e.g., grasses or phytoplankton.

\subsubsection{Ensemble and parameter distributions}

To analyze this model in the thermodynamic limit where the number of species becomes large, we work with random interaction matrices. This approach, inspired by the statistical physics of disordered systems, allows us to identify universal behaviors that are independent of the specific realization of interactions.

We choose to scale the means and variances of the consumer preferences with the number of species, $1/M_N$ or $1/M_R$. This scaling greatly simplifies the mathematical treatment in the thermodynamic limit and ensures that extensive quantities remain well-defined as system size increases.

The predator-consumer interaction matrix is parameterized as:
\begin{align}
d_{\alpha j} &= \frac{\mu_d}{M_N} + \frac{\sigma_d}{\sqrt{M_N}} \gamma_{\alpha j},
&\quad \langle \gamma_{\alpha j} \gamma_{\beta i} \rangle &= \delta_{\alpha \beta} \delta_{ji}, \quad \langle \gamma_{\alpha j} \rangle = 0
\end{align}
where $\mu_d$ sets the mean interaction strength, $\sigma_d$ controls the variance, and $\gamma_{\alpha j}$ are independent standard random variables.

The predation impact matrix, which describes how predators affect consumers, is constructed to have tunable correlation with the benefit matrix:
\begin{align}
e_{\beta i} &= r_2^{-1} \frac{\mu_e}{M_N} + \sqrt{r_2^{-1}} \frac{\sigma_e}{\sqrt{M_N}} 
\left( \rho_X \gamma_{\beta i} + \sqrt{1 - \rho_X^2} \lambda_{\beta i} \right),
&\quad \langle \lambda_{\beta i} \lambda_{\alpha j} \rangle &= \delta_{\beta \alpha} \delta_{ij}, \quad \langle \lambda_{\beta i} \rangle = 0
\end{align}
where $\lambda_{\beta i}$ is an independent zero mean, unit variance random matrix and $\rho_X$ is the predator reciprocity parameter that we will discuss below.

Similarly, the consumer-resource interactions are parameterized as:
\begin{align}
c_{i\alpha} &= \frac{\mu_c}{M_R} + \frac{\sigma_c}{\sqrt{M_R}} \xi_{i\alpha},
&\quad \langle \xi_{i\alpha} \xi_{j\beta} \rangle &= \delta_{ij} \delta_{\alpha \beta}, \quad \langle \xi_{i\alpha} \rangle = 0
\end{align}

\begin{align}
f_{jp} &= r_1^{-1} \frac{\mu_f}{M_R} + \sqrt{r_1^{-1}} \frac{\sigma_f}{\sqrt{M_R}} 
\left( \rho_R \xi_{jp} + \sqrt{1 - \rho_R^2} \zeta_{jp} \right),
&\quad \langle \zeta_{jp} \zeta_{iq} \rangle &= \delta_{ji} \delta_{pq}, \quad \langle \zeta_{jp} \rangle = 0
\end{align}

The mortality rates and carrying capacities are drawn from distributions with small variance around their means:
\begin{align}
u_\alpha &= u + \delta u_\alpha,
& \langle \delta u_\alpha \rangle &= 0,
& \langle \delta u_\alpha \delta u_\beta \rangle &= \sigma_u^2 \delta_{\alpha \beta} 
\end{align}

\begin{align}
m_i &= m + \delta m_i,
& \langle \delta m_i \rangle &= 0,
& \langle \delta m_i \delta m_j \rangle &= \sigma_m^2 \delta_{ij} 
\end{align}

\begin{align}
\kappa_p &= \kappa + \delta \kappa_p,
& \langle \delta \kappa_p \rangle &= 0,
& \langle \delta \kappa_p \delta \kappa_q \rangle &= \sigma_\kappa^2 \delta_{pq} 
\end{align}

We define the ratios of species numbers at adjacent trophic levels:
\begin{equation}
r_1 = \frac{M_N}{M_R}, \qquad r_2 = \frac{M_X}{M_N}
\end{equation}

\subsubsection{Parameter values used in simulations}
Throughout the main text of the manuscript, unless otherwise specified, we used the following parameter values while performing simulations of the model:
\begin{equation}
\begin{aligned}
\mu_c = \mu_f = \mu_d = \mu_e &= 100 \\
u = m = \kappa &= 1 \\
\sigma_c = \sigma_f = \sigma_d = \sigma_e &= 4 \\
\sigma_u = \sigma_m = \sigma_\kappa &= 0.1 \\
M_X = M_N = M_R &= 100\\
\eta=1
\end{aligned}
\end{equation}

We integrated the dynamical equations using SciPy’s \texttt{solve\_ivp} with the LSODA method. We added a small immigration rate $\lambda$ to regularize extinctions, set to $\lambda=10^{-9}$ throughout this work.

\subsubsection{Non-reciprocal interactions}

A key feature of our model is the inclusion of non-reciprocal interactions between trophic levels. In classical consumer-resource models, the benefit a consumer derives from a resource is often assumed to be proportional to the impact on that resource. However, in real ecosystems, these quantities can be decoupled---a species may benefit greatly from a resource while having relatively little impact on it, or vice versa.

We quantify the degree of reciprocity using correlation parameters. The correlation between two random variables $x$ and $y$ is defined as:
\begin{equation}
\mathrm{corr}(x,y) = \frac{\mathrm{cov}(x,y)}{\sqrt{\mathrm{var}(x)\,\mathrm{var}(y)}},
\end{equation}
where $\mathrm{cov}$ represents covariance between two random variables, and $\mathrm{var}$ the variance of a random variable respectively.

The \emph{predator reciprocity} parameter $\rho_X$ controls the correlation between predator benefits and predation impacts:
\begin{equation}
\mathrm{corr}(d_{\alpha j}, e_{\beta i}) = \rho_X \,\delta_{\alpha\beta}\delta_{ji}.
\label{eq:predator_reciprocity}
\end{equation}
When $\rho_X = 1$, the interactions are fully reciprocal---predators that benefit most from a consumer also have the greatest impact on that consumer. When $\rho_X = 0$, benefits and impacts are completely uncorrelated, representing maximally non-reciprocal interactions.

Similarly, the \emph{resource reciprocity} parameter $\rho_R$ controls the correlation between consumer benefits and resource depletion:
\begin{equation}
\mathrm{corr}(c_{iq}, f_{jp}) = \rho_R \,\delta_{ij}\delta_{qp}.
\label{eq:producer_reciprocity}
\end{equation}

\section*{Appendix B: Cavity calculation}
\label{sec:cavity}

\subsubsection{Overview of the cavity method}

The cavity method is a powerful technique from the statistical physics of disordered systems that allows us to compute self-consistent mean-field equations for the steady-state behavior of large, random systems. The key insight is that in the thermodynamic limit, the system exhibits self-averaging behavior: the distribution of properties across constituents becomes independent of the specific realization of quenched disorder. The basic procedure, which we will follow below, consists of the following steps:
\begin{enumerate}
    \item Begin with an ecosystem at steady state containing $M_X$ predators, $M_N$ consumers, and $M_R$ resources.
    \item Add one new ``cavity'' species to each trophic level: a predator $X_0$, a consumer $N_0$, and a resource $R_0$.
    \item Treat the effect of these new species on the existing ecosystem as small perturbations.
    \item Use linear response theory to express how existing species abundances change due to the cavity species.
    \item Exploit the central limit theorem to derive self-consistency equations for the cavity species.
    \item By self-averaging, these equations describe the typical behavior of any species in the ecosystem.
\end{enumerate}

\subsubsection{Steady state before adding cavity species}

Before introducing the cavity species, we assume an ecosystem  at steady state with $M_X$ predator species, $M_N$ consumer species, and $M_R$ resource species. The steady-state conditions are:
\begin{align}
0 &= X_{\alpha\wo0}^* \left( \eta\sum_{j=1}^{M_N} d_{\alpha j} N_{j\wo0}^* - u_\alpha \right), \quad \alpha = 1, \ldots, M_X \\
0 &= N_{i\wo0}^* \left( \eta\sum_{q=1}^{M_R} c_{iq} R_{q\wo0}^* - m_i - \sum_{\beta=1}^{M_X} e_{\beta i} X_{\beta\wo0}^* \right), \quad i = 1, \ldots, M_N \\
0 &= R_{p\wo0}^* \left( \kappa_p - R_{p\wo0}^* - \sum_{j=1}^{M_N} f_{jp} N_{j\wo0}^* \right), \quad p = 1, \ldots, M_R
\end{align}
We denote steady-state abundances with a superscript ${}^*$, though for convenience, going forward we will suppress this notation when the context is clear. For surviving species (those with nonzero abundances), the terms in parentheses must vanish, while for extinct species, the parenthetical term must be negative. The latter implies assuming that the comm1 is at an uninvadable steady-state, where no extinct species can grow when rare.

\subsubsection{Adding cavity species}

We introduce cavity species $X_0$, $N_0$, and $R_0$ to the ecosystem. The steady-state conditions for these cavity species are:

\begin{align}
0 &= X_0^* \left( \eta\sum_j d_{0 j} N_j^* + \eta d_{0 0} N_0^* - u_0 \right) 
\end{align}

\begin{align}
0 &= N_0^* \left( \eta\sum_q c_{0q} R_q^* + \eta c_{00} R_0^* - m_0 
- \sum_\beta e_{\beta 0} X_\beta^* - e_{00} X_0^* \right) 
\end{align}

\begin{align}
0 &= R_0^* \left( \kappa_0 - R_0^* - \sum_j f_{j 0} N_j^* - f_{00} N_0^* \right).
\end{align}
Similarly, the steady-state conditions for the existing species after adding the cavity species become:

\begin{align}
0 &= X_\alpha \left( \eta\sum_j d_{\alpha j} N_j + \eta d_{\alpha 0} N_0 - u_\alpha \right) \\
0 &= N_i \left( \eta\sum_q c_{iq} R_q + \eta c_{i0} R_0 - m_i - \sum_\beta e_{\beta i} X_\beta - e_{0i} X_0 \right) \\
0 &= R_p \left( \kappa_p - R_p - \sum_j f_{jp} N_j - f_{0p} N_0 \right)
\end{align}

\subsubsection{Perturbative treatment}

The new terms introduced by the cavity species can be treated as small perturbations to the original system parameters. Specifically, we identify the effective parameter shifts:
\begin{equation}
\begin{aligned}
u_\alpha &\rightarrow u_\alpha - \eta d_{\alpha 0} N_0 \\
m_i &\rightarrow m_i - \eta c_{i0} R_0 + e_{0i} X_0 \\
\kappa_p &\rightarrow \kappa_p - f_{0p} N_0
\end{aligned}
\end{equation}

In the thermodynamic limit, the effect of a single cavity species on the rest of the ecosystem is of order $O(M^{-1/2})$, which is small. This allows us to use linear response theory to express how species abundances change in response to these perturbations. We expand the perturbed abundances using susceptibility matrices, given by:

\begin{align}
X_\beta = X_{\beta \wo0} 
& - \sum_\alpha \left( \frac{\partial X_\beta}{\partial u_\alpha} \right) \eta d_{\alpha 0} N_0
+ \sum_i \left( \frac{\partial X_\beta}{\partial m_i} \right) \left( e_{0i} X_0 - \eta c_{i0} R_0 \right)
- \sum_p \left( \frac{\partial X_\beta}{\partial \kappa_p} \right) f_{0p} N_0 
\end{align}

\begin{align}
N_j = N_{j\wo0} 
& - \sum_\alpha \left( \frac{\partial N_j}{\partial u_\alpha} \right) \eta d_{\alpha 0} N_0
+ \sum_i \left( \frac{\partial N_j}{\partial m_i} \right) \left( e_{0i} X_0 - \eta c_{i0} R_0 \right)
- \sum_p \left( \frac{\partial N_j}{\partial \kappa_p} \right) f_{0p} N_0 
\end{align}

\begin{align}
R_q = R_{q \wo0} 
& - \sum_\alpha \left( \frac{\partial R_q}{\partial u_\alpha} \right) \eta d_{\alpha 0} N_0
+ \sum_i \left( \frac{\partial R_q}{\partial m_i} \right) \left( e_{0i} X_0 - \eta c_{i0} R_0 \right)
- \sum_p \left( \frac{\partial R_q}{\partial \kappa_p} \right) f_{0p} N_0 
\end{align}

Here, the notation $X_{\beta\wo0}$ denotes the steady-state abundance before adding the cavity species. We define the response functions (susceptibilities) as:
\begin{equation}
\chi^{(X)}_{\beta\alpha} \equiv \frac{\partial X_\beta}{\partial u_\alpha}, \qquad
\nu^X_{\beta i} \equiv \frac{\partial X_\beta}{\partial m_i}, \qquad
K^X_{\beta p} \equiv \frac{\partial X_\beta}{\partial \kappa_p}
\end{equation}

\begin{equation}
\chi^N_{j\alpha} \equiv \frac{\partial N_j}{\partial u_\alpha}, \qquad
\nu^{(N)}_{ji} \equiv \frac{\partial N_j}{\partial m_i}, \qquad
K^N_{jp} \equiv \frac{\partial N_j}{\partial \kappa_p}
\end{equation}

\begin{equation}
\chi^R_{q\alpha} \equiv \frac{\partial R_q}{\partial u_\alpha}, \qquad
\nu^R_{q i} \equiv \frac{\partial R_q}{\partial m_i}, \qquad
K^{(R)}_{q p} \equiv \frac{\partial R_q}{\partial \kappa_p}
\end{equation}

\subsubsection{Top Layer: Predators}

We now derive the self-consistency equation for predators. Substituting the perturbative expansion into the steady-state equation for the cavity predator yields:

\begin{align}
X_0 \bigg( 
& \sum_j \eta d_{0j} N_{j\wo0} 
- \sum_{j\alpha} \eta^2 d_{0j} \chi^N_{j\alpha}  d_{\alpha 0} N_0 
+ \sum_{ji} \eta d_{0j} \nu^{(N)}_{ji} e_{0i} X_0 
- \sum_{ji} \eta^2 d_{0j} \nu^{(N)}_{ji} c_{i0} R_0 \notag \\
& - \sum_{jp} \eta d_{0j} K^N_{jp} f_{0p} N_0
+ \eta d_{00} N_0 - u_0 
\bigg) = 0
\end{align}

We now analyze each term systematically, keeping only those that contribute at leading order in the thermodynamic limit.

\textbf{Term (i):} The first term involves a sum over consumers:
\begin{equation}\sum_j \eta d_{0j} N_{j\wo0} 
 = \sum_j \eta \left( \frac{\mu_d}{M_N} + \frac{\sigma_d}{\sqrt{M_N}} \gamma_{0j} \right) N_{j\wo0} 
 =\eta \mu_d \langle N \rangle + \eta \sum_j \frac{\sigma_d}{\sqrt{M_N}} \gamma_{0j} N_{j\wo0}\end{equation}

The first part gives the mean contribution, while the second part involves a sum of random variables that will contribute to the variance.

\textbf{Term (ii):} The second term involves products of random matrices:
\begin{equation}\sum_{j\alpha} \eta^2
\left( \frac{\mu_d}{M_N} + \frac{\sigma_d}{\sqrt{M_N}} \gamma_{0j} \right) 
\left( \frac{\mu_d}{M_N} + \frac{\sigma_d}{\sqrt{M_N}} \gamma_{\alpha 0} \right) 
\chi^N_{j\alpha} N_0 \simeq 0\end{equation}
This term vanishes at leading order because the dominant contributions scale as $O(M_N^{-1})$.

\textbf{Term (iii):} This term involves the correlation between $d$ and $e$ matrices:
\begin{align*}
&\sum_{ji} \eta \left( \frac{\mu_d}{M_N} + \frac{\sigma_d}{\sqrt{M_N}} \gamma_{0j} \right)
\left( r_2^{-1} \frac{\mu_e}{M_N} + \frac{{\sqrt{r_2^{-1}}\sigma_e}}{\sqrt{M_N}} (\rho_X \gamma_{0i} + \sqrt{1 - \rho_X^2} {\lambda}_{0i}) \right) 
\nu^{(N)}_{ji} X_0 \\
&= \eta \sum_{ji} \frac{\sigma_d \sigma_e}{M_N} \sqrt{r_2^{-1}} \rho_X \gamma_{0j} \gamma_{0i} \nu^{(N)}_{ji} X_0\\
&=\eta \sum_{i} \frac{\sigma_d \sigma_e}{M_N} \sqrt{r_2^{-1}} \rho_X  \nu^{(N)}_{ii} X_0\\
&=  \eta \sigma_d \sigma_e \sqrt{r_2^{-1}} \rho_X \nu^{(N)} X_0
\end{align*}

This is the crucial term where non-reciprocity enters. The correlation $\langle \gamma_{0j} \gamma_{0i} \rangle = \delta_{ji}$ selects out the diagonal susceptibility, and the reciprocity parameter $\rho_X$ modulates the strength. We define the average self-susceptibility:
\begin{equation}
\nu^{(N)} = \frac{1}{M_N} \sum_i \nu^{(N)}_{ii} = \frac{1}{M_N} \operatorname{Tr}(\nu^{(N)})
\end{equation}

\textbf{Term (iv):} This term vanishes because $\gamma$ and $\xi$ are independent, that is:
\begin{equation}\sum_{ji} \eta^2 d_{0j} \nu^{(N)}_{ji} c_{i0} R_0
= \sum_{ji} \eta^2 \left( \frac{\mu_d}{M_N} + \frac{\sigma_d}{\sqrt{M_N}} \gamma_{0j} \right)
\left( \frac{\mu_c}{M_R} + \frac{\sigma_c}{\sqrt{M_R}} \xi_{i0} \right)
\nu^{(N)}_{ji} R_0 \simeq 0\end{equation}

\textbf{Term (v):} Similarly, this term vanishes due to independence of $\gamma$ with $\xi$ and $\zeta$:
\begin{equation}\sum_{jp} \eta d_{0j} K^N_{jp} f_{0p} N_0 = 0\end{equation}

\textbf{Term (vi):} The self-interaction term is negligible:
\begin{equation}\eta d_{00} N_0 \simeq 0 \qquad \left( \text{as } \mathcal{O}\left( \frac{1}{M_N} \right) \right)\end{equation}

\textbf{Term (vii):} The mortality rate decomposes as:
\begin{equation} u_0 = u + \delta u_0 \end{equation}

Combining all terms, the steady-state condition becomes:
\begin{equation}
X_0 \left( \eta \mu_d \langle N \rangle + \eta \frac{\sigma_d}{\sqrt{M_N}} \sum_j \gamma_{0j} N_{j\wo0}
+ \eta \sigma_d \sigma_e \sqrt{r_2^{-1}} \rho_X \nu^{(N)} X_0 - u - \delta u_0 \right) = 0
\end{equation}

By the central limit theorem, the sum of fluctuating terms converges to a Gaussian:
\begin{equation}
\eta \frac{\sigma_d}{\sqrt{M_N}} \sum_j \gamma_{0j} N_{j\wo0} - \delta u_0 = \sqrt{\sigma_u^2 + \eta^2 \sigma_d^2 \langle N^2 \rangle} \ Z_X,
\qquad \text{where } Z_X \sim \mathcal{N}(0,1)
\end{equation}

Solving for $X_0$ and keeping only non-negative solutions (discarding invadable states):

\begin{equation}
X_0 = \max\left( 0,  \frac{ \eta \mu_d \langle N \rangle + \sqrt{ \sigma_u^2 + \eta^2 \sigma_d^2 \langle N^2 \rangle } \ Z_X - u }
{ -\eta \sigma_d \sigma_e \rho_X \nu^{(N)} \sqrt{r_2^{-1}} } \right)
\end{equation}

This expression shows that predator abundance is a truncated Gaussian random variable. The numerator captures the balance between benefits from predation and mortality costs, while the denominator involves the self-susceptibility $\nu^{(N)}$ modulated by the reciprocity parameter $\rho_X$.

\subsubsection{Middle Layer: Consumers}

We repeat the analysis for consumers, which experience both top-down pressure from predators and bottom-up benefits from resources. Substituting the perturbative expansions:

\begin{align}
N_0 \Bigg[
& \sum_q \eta c_{0q} R_{q\wo0}
- \sum_{q\alpha} \eta^2 c_{0q} \chi^R_{q\alpha} d_{\alpha 0} N_0
+ \sum_{qi} \eta c_{0q} \nu^R_{qi} e_{0i} X_0
- \sum_{qi} \eta c_{0q} \nu^R_{qi} c_{i0} R_0 - \sum_{qp} \eta c_{0q} K^{(R)}_{qp} f_{0p} N_0 \notag \\
& + \eta c_{00} R_0 - m_0 - \sum_\beta e_{\beta 0} X_{\beta\wo0} \notag  + \sum_{\beta\alpha} \eta e_{\beta 0} \chi^{(X)}_{\beta \alpha} d_{\alpha 0} N_0
- \sum_{\beta i} e_{\beta 0} \nu^X_{\beta i} e_{0i} X_0
+ \sum_{\beta i} \eta e_{\beta 0} \nu^X_{\beta i} c_{i0} R_0 \notag \\
& + \sum_{\beta p} e_{\beta 0} K^X_{\beta p} f_{0p} N_0
- e_{00} X_0
\Bigg] = 0
\end{align}

Analyzing each term systematically:

\textbf{Term (i):} Resource benefits contribute:
\begin{equation}\sum_q \eta c_{0q} R_{q\wo0}
= \eta \mu_c \langle R \rangle + \sum_q \eta \frac{\sigma_c}{\sqrt{M_R}} \xi_{0q} R_{q\wo0}\end{equation}

\textbf{Terms (ii)-(iv):} These vanish due to independence of random matrices.

\textbf{Term (v):} The resource-consumer correlation term:
\begin{equation}
\sum_{qp} \eta c_{0q} {K}^{R}_{qp} f_{0p} N_0
= \rho_R \eta {\sigma_c \sigma_f \sqrt{r_1^{-1}}}{{K^{(R)}}} N_0
\end{equation}
where we define $K^{(R)} = \frac{1}{M_R} \operatorname{Tr}({K}^{R})$.

\textbf{Term (viii):} Predation pressure:
\begin{equation}\sum_\beta e_{\beta 0} X_{\beta\wo0} 
= \mu_e \langle X\rangle + \sum_\beta \frac{\sigma_e \sqrt{r_2^{-1}}}{\sqrt{M_N}} 
\rho_X \gamma_{\beta 0} X_{\beta \wo0} +\sum_\beta \frac{\sigma_e \sqrt{r_2^{-1}}}{\sqrt{M_N}} 
\sqrt{1-{\rho_X^2}} \lambda_{\beta 0} X_{\beta \wo0}\end{equation}

\textbf{Term (ix):} The predator susceptibility term:
\begin{equation}
\sum_{\beta \alpha}\eta  e_{\beta 0} \chi^{(X)}_{\beta \alpha} d_{\alpha 0} N_0
=  \eta \sigma_d \sigma_e \sqrt{r_2^{-1}} \rho_X \chi^{(X)}  N_0
\end{equation}
where we define $\chi^{(X)} = \frac{1}{M_X} \operatorname{Tr}({\chi}^{X})$.

\textbf{Terms (x)-(xiii):} These vanish or are negligible.

Combining and applying the central limit theorem:

\begin{align*}
0 = {} & 
N_0^{*} \left( 
    \eta \mu_c \langle R \rangle 
    + \sum_{q} \frac{\eta \sigma_c}{\sqrt{M_R}} \, \xi_{0q} R_{q\wo0}^{*} 
    - \eta \sigma_c \sigma_f K^{(R)}\sqrt{r_1^{-1}}\rho_R \, N_0
    - m - \delta m_0 - \mu_e \langle X \rangle \right.
    \\
& \left.
    - \sum_{\beta} 
        \frac{\sigma_e \sqrt{r_2^{-1}}  }{\sqrt{M_N}}  \rho_X\gamma_{\beta 0} X_{\beta/ 0}^{*}- \sum_{\beta} 
        \frac{\sigma_e}{\sqrt{M_N}}
        \sqrt{r_2^{-1}} \sqrt{1 - \rho_X^{2}} \,
        \lambda_{\beta 0} X_{\beta\wo0}^{*}
    + \eta \rho_X r_2 \sigma_e \sigma_d  \chi^{(X)} N_0^{*} \sqrt{r_2^{-1}}
\right)
\end{align*}

The fluctuating terms combine into a Gaussian:
\begin{equation}\sqrt{\sigma_m^{2} + \eta^2 \sigma_c^{2} \langle R^{2} \rangle + \sigma_e^{2} \langle X^{2} \rangle}\; Z_{N},
    \quad\text{where } Z_N \sim \mathcal{N}(0,1)\end{equation}

The final expression for consumer abundance:
\begin{equation}
N_0^{*} = \max \left(
0,\;
\frac{
    \eta \mu_c \langle R \rangle 
    - m 
    - \mu_e \langle X \rangle
    + \sqrt{\sigma_m^{2} + \eta^2 \sigma_c^{2} \langle R^{2} \rangle + \sigma_e^{2} \langle X^{2} \rangle}\; Z_{N}
}{
    \eta \sigma_c \sigma_f \sqrt{r_1^{-1}} \rho_R K^{(R)}
    - \eta \sigma_d \sigma_e \sqrt{r_2^{-1}} \rho_X r_2 \chi^{(X)}
}
\right)
\end{equation}

Note that the denominator contains contributions from both reciprocity parameters, reflecting the consumer's intermediate position in the trophic structure.

\subsubsection{Bottom Layer: Resources}

For resources, the analysis is simpler as they only experience top-down pressure from consumers:

\begin{align}
0 = {} & R_0^{*} \Bigg(
    \kappa_0 - R_0^{*}
    - \sum_{j} f_{j0} N_{j\wo0}^{*}
    + \sum_{j\alpha} \eta f_{j0} {\chi}^{N}_{j\alpha} d_{\alpha 0} N_0^{*} 
    - \sum_{ji} f_{j0} \nu_{ji}^{N} \, e_{0i} X_{0}^{*} \notag \\
    &+ \sum_{ji} \eta f_{j0} \nu_{ji}^{N} c_{i0} R_{0}^{*} 
   + \sum_{jp} f_{j0} K_{jp}^{N} f_{0p} N_0^{*}
    - f_{00} N_0^{*}
\Bigg)
\end{align}

The key terms are:

\textbf{Term (iii):} Consumer impact on resources:
\begin{align*}
\sum_{j} f_{j0} N_{j\wo0} 
&= \mu_f \langle N \rangle
    + \sum_{j} \frac{\sigma_f}{\sqrt{M_N}} \rho_R \xi_{j0} N_{j\wo0}
    + \sum_{j} \frac{\sigma_f}{\sqrt{M_N}} \sqrt{1 - \rho_R^2} \, \zeta_{j0} N_{j\wo0}
\end{align*}

\textbf{Term (vi):} The consumer-resource correlation term:
\begin{align*}
\sum_{ji} \eta f_{j0} \nu_{ji}^N c_{i0}R_0
&= \eta \sqrt{ r_1}\, \sigma_f \sigma_c \rho_R \nu^{(N)} R_0
\end{align*}

The steady-state condition becomes:
\begin{align*}
\kappa + \delta \kappa_0 - R_0 - \mu_f \langle N \rangle 
- \sum_j\frac{ \sigma_f }{\sqrt{M_N}}{\rho_R}\,\xi_{j0} N_{j\wo0}
- \sum_j\frac{ \sigma_f}{\sqrt{M_N}} \sqrt{1-\rho_R^2}\,\zeta_{j0} N_{j\wo0} 
+ \eta \sqrt{ r_1} \,\sigma_f \sigma_c \rho_R \nu^{(N)} R_0 = 0 
\end{align*}

Applying the central limit theorem to the fluctuating terms, we get:
\begin{equation}\sqrt{\sigma_\kappa^2 + \sigma_f^2 \langle N^2 \rangle} {Z}_{R},\quad\text{where } Z_R \sim \mathcal{N}(0,1)\end{equation}

Thus, solving the resource mean-field equation, we obtain the following expression for the steady-state cavity resource abundance:
\begin{equation}
R_0 = \max \left( 0,\; 
\frac{\kappa - \mu_f \langle N \rangle + \sqrt{\sigma_\kappa^2 + \sigma_f^2 \langle N^2 \rangle} {Z}_{R}}
{1 - \eta \sqrt{ r_1}\, \sigma_f \sigma_c \rho_R \nu^{(N)} } \right)
\end{equation}

\subsubsection{Summary of cavity solutions}

The cavity calculation yields steady-state abundances for each trophic level that take the form of truncated Gaussian random variables. This structure has a natural physical interpretation which we now describe.

Each cavity species interacts with $O(M)$ other species through random interaction coefficients. By the central limit theorem, the net effect of these many random contributions converges to a Gaussian distribution in the large-$M$ limit. For instance, the growth rate of the cavity predator $X_0$ depends on the sum $\sum_j d_{0j} N_j$ over all consumers---a sum of many random variables that becomes normally distributed. The mean of each Gaussian reflects the typical environment experienced by a cavity species, while the variance captures fluctuations arising from the quenched disorder in the interaction coefficients.

The $\max\{0, \cdot\}$ operation enforces the biological constraint that species abundances cannot be negative. Species whose would-be abundance falls below zero are driven to extinction; only those with positive abundance survive. This truncation divides species into two classes: survivors (with $X_0 > 0$, $N_0 > 0$, or $R_0 > 0$) and extinct species. The survival fractions $\phi_X$, $\phi_N$, and $\phi_R$ emerge naturally as the probabilities that the corresponding Gaussian exceeds zero.

The denominators in each expression involve susceptibilities ($\chi^{(X)}$, $\nu^{(N)}$, $K^{(R)}$) and reciprocity parameters ($\rho_X$, $\rho_R$). These encode how the ecosystem responds to the cavity species and, crucially, how that response feeds back onto the cavity species itself. When interactions are reciprocal ($\rho \to 1$), the feedback is strong and the effective self-interaction in the denominator is large, suppressing abundance fluctuations. As reciprocity decreases ($\rho \to 0$), this feedback weakens, and the system becomes increasingly susceptible to perturbations.

\paragraph{Summary of cavity solutions.}
The steady-state abundances are:
\begin{align}
X_0 &= \max\left\{ 0, \frac{\eta \mu_d \langle N \rangle - u + \sqrt{\sigma_u^2 + \eta^2 \sigma_d^2 \langle N^2 \rangle} \, Z_X}{-\eta \sigma_d \sigma_e \rho_X \nu^{(N)} \sqrt{r_2^{-1}}} \right\} \\[6pt]
N_0 &= \max\left\{ 0, \frac{\eta \mu_c \langle R \rangle - m - \mu_e \langle X \rangle + \sqrt{\sigma_m^2 + \eta^2 \sigma_c^2 \langle R^2 \rangle + \sigma_e^2 \langle X^2 \rangle} \, Z_N}{\eta \sigma_c \sigma_f \sqrt{r_1^{-1}} \rho_R K^{(R)} - \eta \sigma_d \sigma_e \sqrt{r_2^{-1}} \rho_X r_2 \chi^{(X)}} \right\} \\[6pt]
R_0 &= \max\left\{ 0, \frac{\kappa - \mu_f \langle N \rangle + \sqrt{\sigma_\kappa^2 + \sigma_f^2 \langle N^2 \rangle} \, Z_R}{1 - \eta \sqrt{r_1} \sigma_f \sigma_c \rho_R \nu^{(N)}} \right\}
\end{align}
where $Z_X$, $Z_N$, $Z_R \sim \mathcal{N}(0,1)$ are independent standard normal random variables. Each expression has the structure:
\begin{equation}
\text{abundance} = \max\left\{ 0, \frac{\text{(mean growth rate)} + \text{(fluctuations)} \times Z}{\text{(feedback term)}} \right\}.
\end{equation}

The numerator determines whether a species can invade: a positive mean growth rate (e.g., $\eta \mu_d \langle N \rangle > u$ for predators) indicates favorable average conditions, while the noise term allows some species to survive even when mean conditions are unfavorable (or go extinct when conditions are favorable). The denominator regularizes the solution through self-consistent feedback from the rest of the ecosystem.

\section*{Appendix C: Self-consistency equations}
\label{sec:selfconsistency}
\subsubsection{Moments of truncated Gaussians}

The cavity solutions express abundances as truncated Gaussian random variables. For a truncated Gaussian of the form:
\begin{equation}
Y = \max \left(\, 0,\, \frac{a}{b} + \frac{c}{b}Z\, \right),
\quad\text{where } Z\sim \mathcal{N}(0,1)
\end{equation}

we can compute the moments as follows. Define the auxiliary functions:
\begin{equation}
w_0(x) = \frac{1}{2} \left(1 + \operatorname{erf}\left(\frac{x}{\sqrt{2}}\right)\right)
\end{equation}

\begin{equation}
w_1(x) = \frac{1}{\sqrt{2\pi}} e^{-\frac{x^2}{2}} + \frac{x}{2} \left(1 + \operatorname{erf}\left(\frac{x}{\sqrt{2}}\right)\right)
\end{equation}

\begin{equation}
w_2(x) = w_0(x) + x \cdot w_1(x)
\end{equation}

The zeroth moment (survival probability), first moment and second moment, are given by:
\begin{equation}
\langle \Theta(Y) \rangle = w_0\left(\frac{a}{c}\right)
\end{equation}

\begin{equation}
\langle Y \rangle = \frac{c}{b} \cdot w_1\left(\frac{a}{c}\right)
\end{equation}

\begin{equation}
\langle Y^2 \rangle = \left(\frac{c}{b}\right)^2 w_2\left(\frac{a}{c}\right)
\end{equation}
\subsubsection{Complete self-consistency equations}
Using the self-averaging property of the system, the cavity species statistics equal the population statistics. This yields 12 self-consistency equations for 12 unknowns:
\begin{equation}
\phi_X, \; \langle X \rangle, \; \langle X^2 \rangle, \;
\phi_N, \; \langle N \rangle, \; \langle N^2 \rangle, \;
\phi_R, \; \langle R \rangle, \; \langle R^2 \rangle, \;
\chi^{(X)}, \; \nu^{(N)}, \; K^{(R)}
\end{equation}

where $\phi_X = M_X^*/M_X$, $\phi_N = M_N^*/M_N$, and $\phi_R = M_R^*/M_R$ are the fractions of surviving species at each level. Using the definitions for the moments of truncated Gaussians equations above, we obtain these equations as:

\begin{align}
\phi_X &= w_0\left(\frac{\eta \mu_d \cdot \langle N \rangle - u}{\sqrt{\sigma_u^2 + \eta^2 \sigma_d^2 \cdot \langle N^2 \rangle}}\right)\label{eq:fullselfcon1}\\
\langle X \rangle &= \left(\frac{\sqrt{\sigma_u^2 + \eta^2 \sigma_d^2 \cdot \langle N^2 \rangle}}{-\sigma_e \eta \sigma_d \sqrt{r_2^{-1}} \rho_X \nu^{(N)}}\right) w_1\left(\frac{\eta \mu_d \cdot \langle N \rangle - u}{\sqrt{\sigma_u^2 + \eta^2 \sigma_d^2 \cdot \langle N^2 \rangle}}\right) \\
\langle X^2 \rangle &= \left(\frac{\sqrt{\sigma_u^2 + \eta^2 \sigma_d^2 \cdot \langle N^2 \rangle}}{-\sigma_e \eta \sigma_d \sqrt{r_2^{-1}} \rho_X \nu^{(N)}}\right)^2 w_2\left(\frac{\eta \mu_d \cdot \langle N \rangle - u}{\sqrt{\sigma_u^2 + \eta^2 \sigma_d^2 \cdot \langle N^2 \rangle}}\right)\\
\phi_N &= w_0\left(\frac{\eta \mu_c \cdot \langle R \rangle - \mu_e \cdot \langle X \rangle - m}{\sqrt{\sigma_m^2 + \eta^2 \sigma_c^2 \cdot \langle R^2 \rangle + \sigma_e^2 \cdot \langle X^2 \rangle}}\right) \\
\langle N \rangle &= \left(\frac{\sqrt{\sigma_m^2 + \eta^2 \sigma_c^2 \cdot \langle R^2 \rangle + \sigma_e^2 \cdot \langle X^2 \rangle}}{\eta \sigma_c \sigma_f \sqrt{r_1^{-1}} \rho_R K^{(R)} - \eta \sigma_d \sigma_e \sqrt{r_2^{-1}} \rho_X r_2 \chi^{(X)}}\right) w_1\left(\frac{\eta \mu_c \cdot \langle R \rangle - \mu_e \cdot \langle X \rangle - m}{\sqrt{\sigma_m^2 + \eta^2 \sigma_c^2 \cdot \langle R^2 \rangle + \sigma_e^2 \cdot \langle X^2 \rangle}}\right) \\
\langle N^2 \rangle &= \left(\frac{\sqrt{\sigma_m^2 + \eta^2 \sigma_c^2 \cdot \langle R^2 \rangle + \sigma_e^2 \cdot \langle X^2 \rangle}}{\eta \sigma_c \sigma_f \sqrt{r_1^{-1}} \rho_R K^{(R)} - \eta \sigma_d \sigma_e \sqrt{r_2^{-1}} \rho_X r_2 \chi^{(X)}}\right)^2 w_2\left(\frac{\eta \mu_c \cdot \langle R \rangle - \mu_e \cdot \langle X \rangle - m}{\sqrt{\sigma_m^2 + \eta^2 \sigma_c^2 \cdot \langle R^2 \rangle + \sigma_e^2 \cdot \langle X^2 \rangle}}\right) \\
\phi_R &= w_0\left(\frac{-\mu_f \cdot \langle N \rangle + \kappa}{\sqrt{\sigma_\kappa^2 + \sigma_f^2 \cdot \langle N^2 \rangle}}\right) \\
\langle R \rangle &= \frac{\sqrt{\sigma_\kappa^2 + \sigma_f^2 \cdot \langle N^2 \rangle}}{1 - \eta \sigma_c \sigma_f \sqrt{r_1^{-1}} r_1 \rho_R \nu^{(N)}} w_1\left(\frac{-\mu_f \cdot \langle N \rangle + \kappa}{\sqrt{\sigma_\kappa^2 + \sigma_f^2 \cdot \langle N^2 \rangle}}\right) \\
\langle R^2 \rangle &= \left(\frac{\sqrt{\sigma_\kappa^2 + \sigma_f^2 \cdot \langle N^2 \rangle}}{1 - \eta \sigma_c \sigma_f \sqrt{r_1^{-1}} r_1 \rho_R \nu^{(N)}}\right)^2 w_2\left(\frac{-\mu_f \cdot \langle N \rangle + \kappa}{\sqrt{\sigma_\kappa^2 + \sigma_f^2 \cdot \langle N^2 \rangle}}\right)\\
\chi^{(X)} &= \frac{\phi_X}{\sigma_e \eta \sigma_d \sqrt{r_2^{-1}} \rho_X \nu^{(N)}}\label{eq:selfcon1} \\
\nu^{(N)} &= -\frac{\phi_N}{\eta \sigma_c \sigma_f \sqrt{r_1^{-1}} \rho_R K^{(R)} - \eta \sigma_d \sigma_e \sqrt{r_2^{-1}} \rho_X r_2 \chi^{(X)}} \\
K^{(R)} &= \frac{\phi_R}{1 - \eta \sigma_c \sigma_f \sqrt{r_1^{-1}} r_1 \rho_R \nu^{(N)}}–\label{eq:selfcon2}
\end{align}
These 12 equations can be simultaneously solved using numerical nonlinear least squares methods. These analytical solutions agree well with numerical simulations of the full models, as shown using representative examples in Fig.~\ref{fig:s1}.
\begin{figure}[ht!]
    \centering

    \includegraphics[width=\textwidth]{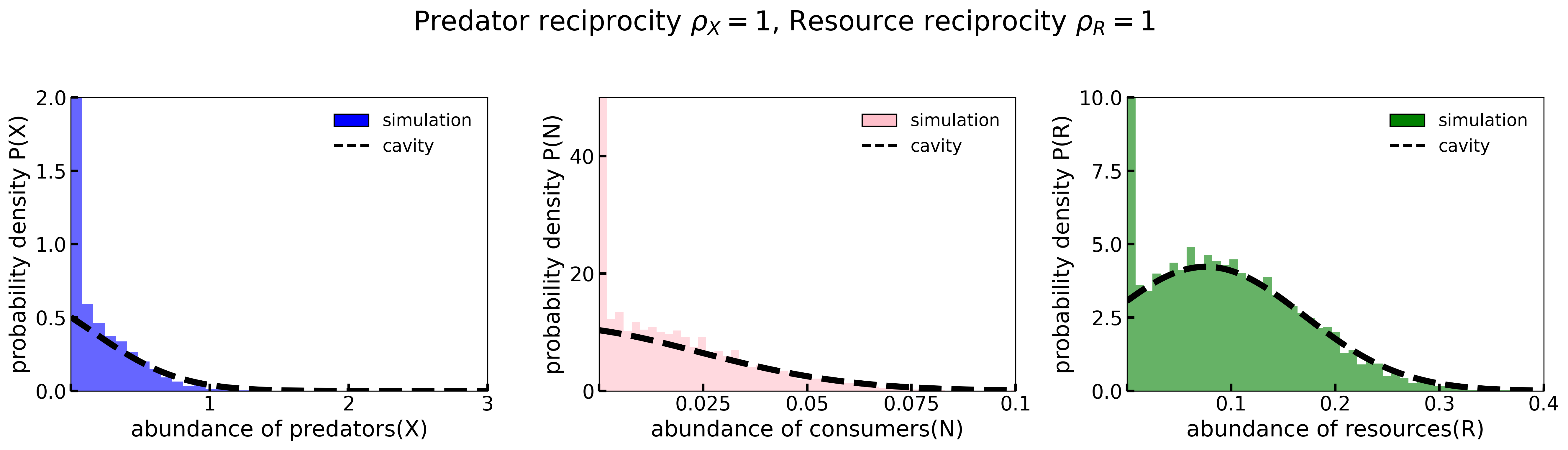}
    \vspace{2mm}

    \includegraphics[width=\textwidth]{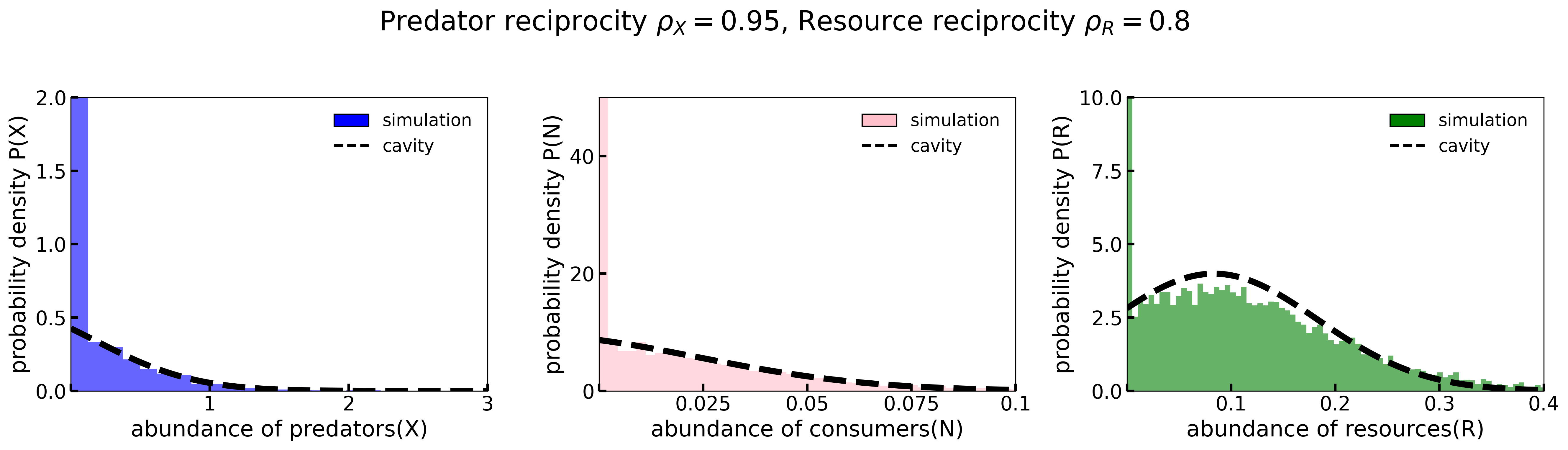}
    \vspace{2mm}

    \includegraphics[width=\textwidth]{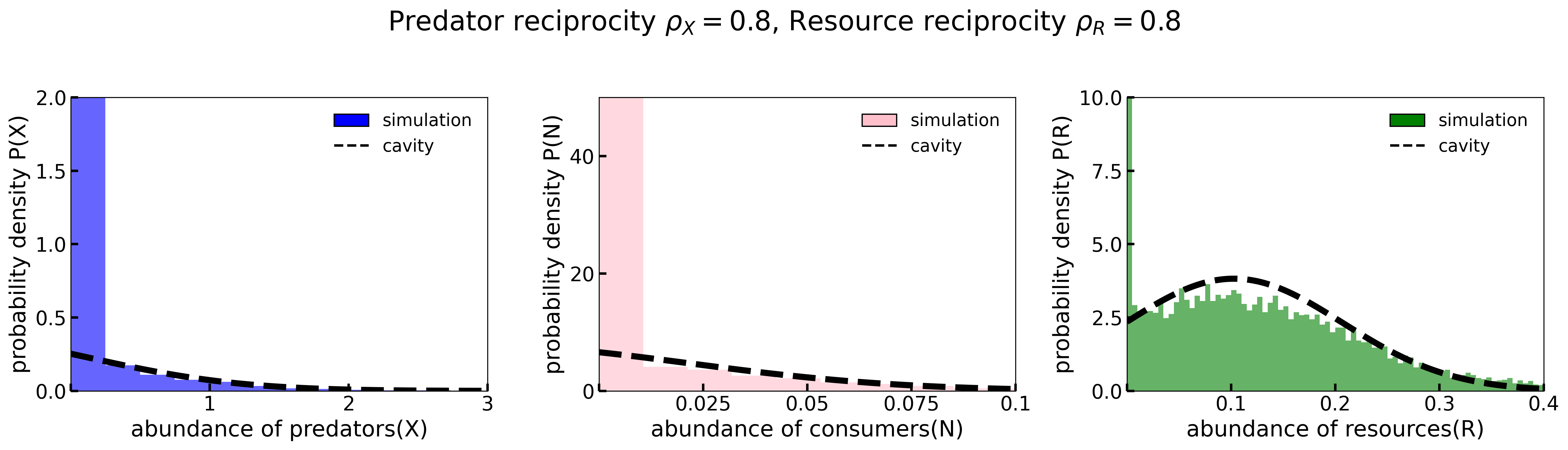}
    \vspace{2mm}

    \includegraphics[width=\textwidth]{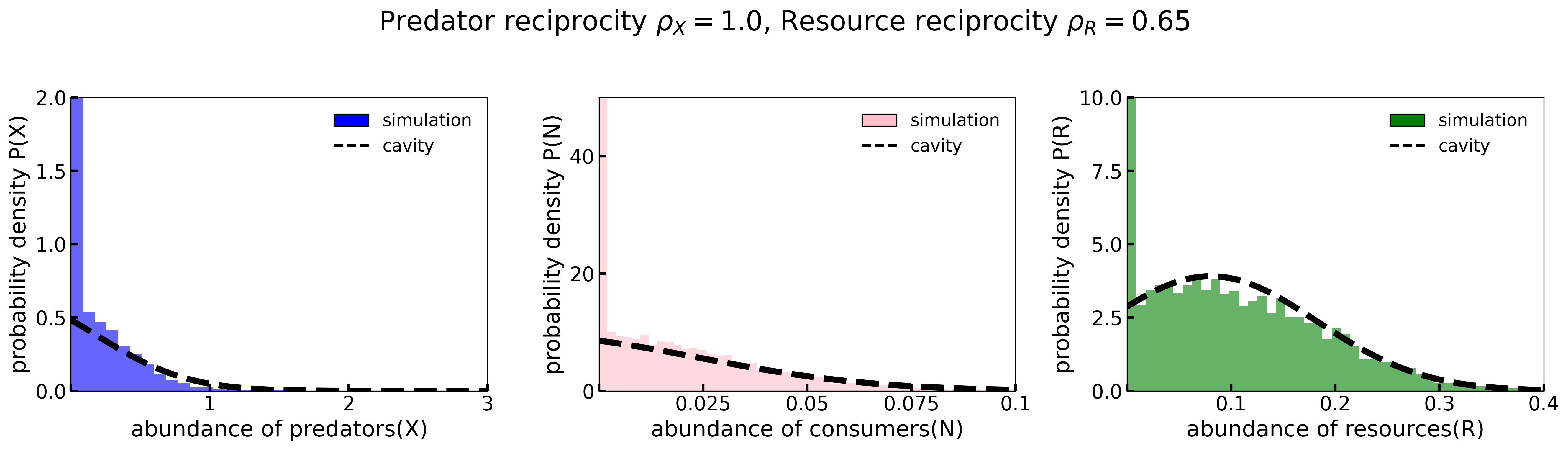}

    \caption{\justifying Steady-state abundance distributions from simulations (100 realizations histograms) and cavity theory predictions (dashed lines). Peaks at zero indicate extinct species. The cavity predictions show good agreement with the simulation results.Panels correspond to different combinations of predator and resource reciprocity
    parameters $(\rho_X, \rho_R)$ as indicated.
    Remarkably, even in the non-reciprocal regime ($\rho_X \neq 1$, $\rho_R \neq 1$),
    the cavity-theory predictions agree with the numerical simulations.
    }
    \label{fig:s1}
\end{figure}\\

\section*{Appendix D: Infeasibility transition and boundary}
\label{sec:infeasibility}

The cavity solution becomes infeasible when susceptibilities diverge. This occurs at specific boundaries in parameter space that have important ecological interpretations. Substituting the susceptibility equations into each other, when $\nu^{(N)} \to 0$, we find from the cavity equations that $\langle X \rangle \to \infty$ (with $\chi^{(X)} \to \infty$ and $K^{(R)} \to \phi_R$). This leads to the condition:
\begin{align}
M_N \phi_N = M_X \phi_X
\label{eq:infeas1}
\end{align}

This can be interpreted as follows. One way for the model to become infeasible is when the the number of surviving consumers equals the number of surviving predators. This represents a boundary where the consumer trophic level cannot support more predators. Similarly, when $\chi^{(X)} \to 0$, we find $\langle N \rangle \to \infty$. This leads to:
\begin{align}
M_N \phi_N = M_R \phi_R + M_X \phi_X
\label{eq:infeas2}
\end{align}

This condition can be interpreted as occurring when the number of surviving consumers equals the sum of surviving resources and predators. This represents the competitive exclusion bound for the consumer trophic level. Note that $\langle R \rangle$ cannot diverge because that would require $K^{(R)} \to \infty$, which is inconsistent with the structure of the equations. Whichever of the two conditions in Eqs.~\eqref{eq:infeas1}--\eqref{eq:infeas2} occurs first will lead to infeasibility. In Fig.~\ref{fig:infeasboundary}, we plot the theoretical infeasibility boundary as a dashed white curve, which we find agrees well with the divergence in the value of the least-squares objective function we use to numerically solve the cavity self-consistency equations Eqs.~\eqref{eq:fullselfcon1}--\eqref{eq:selfcon2}. We find that it is always Eq.~\eqref{eq:infeas1} that occurs at the infeasibility boundary, i.e., $\nu^{(N)}\to0$.

\begin{figure}[ht!]
    \centering
    \includegraphics[width=0.6\linewidth]{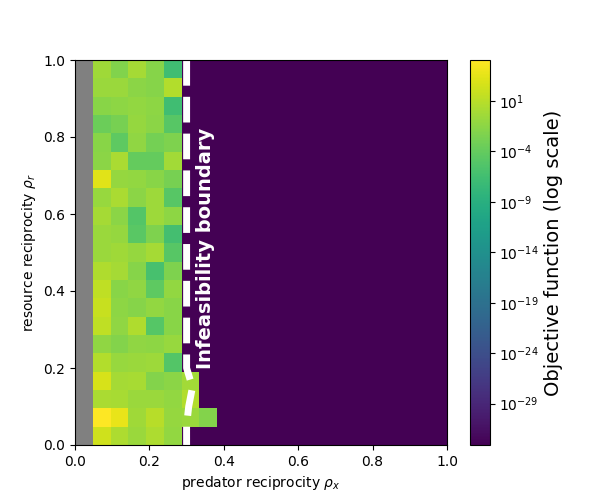}
    \caption{\justifying Least-squares objective values for solving the cavity self-consistency equations. The dashed line indicates the analytically-derived infeasibility boundary (solved numerically using the self-consistency equations and Eqs.~\eqref{eq:infeas1}--\eqref{eq:infeas1}). Regions where the objective function is small correspond to valid solutions of the cavity equations, whereas regions with relatively large objective values represent invalid solutions. The parameter values are the same as mentioned in Appendix~A.}
    \label{fig:infeasboundary}
\end{figure}

\section*{Appendix E: Stability analysis and instability boundary}
\label{sec:instability}

\subsubsection{Perturbing the typical steady state}

To determine when the steady-state solution becomes unstable, we perturb all surviving species at each trophic level (labeled with ${}^+$) by small random amounts and examine how the perturbations propagate through the ecosystem. We perturb the surviving predators, consumers, and resources as:
\begin{equation}
X_{\beta\backslash 0}^+ \to X_{\beta\backslash 0}^+ + \varepsilon \eta_\beta^{(X)}, 
\quad
N_{j\backslash 0}^+ \to N_{j\backslash 0}^+ + \varepsilon \eta_j^{(N)}, 
\quad
R_{\alpha\backslash 0}^+ \to R_{\alpha\backslash 0}^+ + \varepsilon \eta_\alpha^{(R)},
\end{equation}
where $\varepsilon$ is a small parameter and $\eta_i$ are independent random variables with $\langle \eta_i \rangle = 0$ and $\langle \eta_i^2 \rangle = 1$.

\subsubsection{Response to perturbations}

Applying the perturbation to the cavity equations and differentiating with respect to $\varepsilon$, we obtain the response at each trophic level.

For predators:
\begin{equation}
\frac{d X_0^+}{d\varepsilon}
= \frac{\eta \sigma_d}{ - \eta \sigma_d \sigma_e \rho_X \nu^{(N)} \sqrt{r_2^{-1}} \sqrt{M_N}}
\sum_j \gamma_{0j}\left( \frac{d N_j^+}{d\varepsilon}
+ \eta_j^{(N)} \right).
\end{equation}

For consumers, the response involves contributions from both resources below and predators above:
\begin{align}
\frac{dN_0^+}{d\varepsilon}
&=
\frac{1}{
\eta \sigma_c \sigma_f K^{(R)}\rho_R\sqrt{r_1^{-1}}
-
\eta \rho_X\sigma_e\sigma_d\chi^{(X)}\sqrt{r_2}
}
\nonumber\\[6pt]
&\quad \times \Bigg[
\sum_Q \frac{\eta \sigma_c}{\sqrt{M_R}}\xi_{0Q}
\left(\frac{dR_{Q\backslash 0}^+}{d\varepsilon}+\eta_Q^{(R)}\right)
\nonumber\\[4pt]
&\qquad\quad
-
\sum_\beta \frac{\sigma_e \sqrt{r_2^{-1}}}{\sqrt{M_N}}
\left( \rho_X \gamma_{\beta 0} + \sqrt{1-\rho_X^2}\,\lambda_{\beta0} \right)
\left(\frac{dX_{\beta\backslash 0}^+}{d\varepsilon}+\eta_\beta^{(X)}\right)
\Bigg].
\end{align}

For resources:
\begin{equation}
\frac{dR_0^+}{d\varepsilon} =
\frac{- \sum_j \frac{\sigma_f}{\sqrt{M_N}} 
\left( \rho_R \xi_{j0} + \sqrt{1-\rho_R^2} \, \zeta_{j0} \right)
\left(\frac{dN_{j\backslash 0}^+}{d\varepsilon}+\eta_j^{(N)}\right)}
{1 - \eta \sigma_f \sigma_c \rho_R \nu^{(N)} \sqrt{r_1}}.
\end{equation}

\subsubsection{Moments of the response}

These responses are themselves random variables, and thus we are interested in their moments. Due to the symmetric nature of the perturbations, the first moments vanish:
\begin{equation}
\left\langle \frac{d X_0^+}{d\varepsilon} \right\rangle = \left\langle \frac{d N_0^+}{d\varepsilon} \right\rangle = \left\langle \frac{d R_0^+}{d\varepsilon} \right\rangle = 0.
\end{equation}

However, the second moments are nonzero and provide information about stability. Computing these using self-averaging, we obtain three coupled equations for the three second moments as follows:
\begin{equation}
\left\langle \left(\frac{d X_0^+}{d\varepsilon} \right)^2\right\rangle
= \frac{\phi_N}{ \sigma_e^2 \rho_X^2 (\nu^{(N)})^2 r_2^{-1} }
\left(\left\langle \left(\frac{d N_0^+}{d\varepsilon} \right)^2\right\rangle +1\right),
\end{equation}

\begin{align}
\left\langle \left(\frac{d N_0^+}{d\varepsilon} \right)^2\right\rangle
&= \frac{1}{\left(\eta \sigma_c \sigma_f K^{(R)} \rho_R \sqrt{r_1^{-1}} 
- \eta \rho_X \sigma_e \sigma_d \chi^{(X)} \sqrt{r_2}\right)^2 }
\nonumber\\[4pt]
&\quad \times \left[
\eta^2 \sigma_c^2 \phi_R \left(\left\langle \left(\frac{d R_0^+}{d\varepsilon} \right)^2\right\rangle +1\right)
+ \sigma_e^2 \phi_X \left(\left\langle \left(\frac{d X_0^+}{d\varepsilon} \right)^2\right\rangle +1\right)
\right],
\end{align}

\begin{equation}
\left\langle \left(\frac{d R_0^+}{d\varepsilon} \right)^2\right\rangle
= \frac{\sigma_f^2 \phi_N}{ \left(1 - \eta \sigma_f \sigma_c \rho_R \nu^{(N)} \sqrt{r_1}\right)^2}
\left(\left\langle \left(\frac{d N_0^+}{d\varepsilon} \right)^2\right\rangle +1\right).
\end{equation}

\subsubsection{Instability condition}

These coupled equations can be written in matrix form $\mathbf{A}\mathbf{X} = \mathbf{B}$. To express the matrix compactly, we define the effective feedback at the resource and consumer trophic levels:
\begin{align}
\Delta_N &= \eta \sigma_c \sigma_f K^{(R)} \rho_R \sqrt{r_1^{-1}} 
      - \eta \rho_X \sigma_e \sigma_d \chi^{(X)} \sqrt{r_2}, \\[4pt]
\Delta_R &= 1 - \eta \sigma_f \sigma_c \rho_R \nu^{(N)} \sqrt{r_1}.
\end{align}

The matrix equation then takes the form:
\begin{equation}
{
\begin{pmatrix}
\sigma_e^2 \rho_X^2 (\nu^{(N)})^2 r_2^{-1} & -\phi_N & 0 \\[6pt]
-\sigma_e^2 \phi_X & \Delta_N^2 & -\eta^2 \sigma_c^2 \phi_R \\[6pt]
0 & -\sigma_f^2 \phi_N & \Delta_R^2
\end{pmatrix}
\begin{pmatrix}
\left\langle \left( \frac{d X_0^+}{d\varepsilon} \right)^2 \right\rangle \\[8pt]
\left\langle \left( \frac{d N_0^+}{d\varepsilon} \right)^2 \right\rangle \\[8pt]
\left\langle \left( \frac{d R_0^+}{d\varepsilon} \right)^2 \right\rangle
\end{pmatrix}
=
\begin{pmatrix}
\phi_N \\[8pt]
\eta^2 \sigma_c^2 \phi_R + \sigma_e^2 \phi_X \\[8pt]
\sigma_f^2 \phi_N
\end{pmatrix}
}
\end{equation}

Note that $\Delta_N$ captures the competition between bottom-up energy flux (first term) and top-down predation pressure (second term) experienced by consumers, while $\Delta_R$ encodes the balance between intrinsic self-regulation and consumer-induced depletion at the resource level.

The second moments diverge when the determinant of the $3\times3$ matrix $\mathbf{A}$ vanishes:
\begin{equation}
\det(\mathbf{A}) = 0,
\label{eq:instabcondition}
\end{equation}
signaling the breakdown of the replica-symmetric (self-averaging) assumption and the onset of dynamical instability. Setting $\det(\mathbf{A}) = 0$ yields the stability boundary:

\begin{equation}
\begin{aligned}
\left( 1 - \eta \sqrt{r_1}\,\nu^{(N)} \rho_R^\star \sigma_c \sigma_f \right)^2
\Bigg[
&- \sigma_e^2 \, \phi_N \phi_X +\frac{{\rho_X^\star}^2 \sigma_e^2(\nu^{(N)})^2}{r_2}
\left(
\frac{\eta K^{(R)} \sigma_c \sigma_f \rho_R^\star}{\sqrt{r_1}}
-\eta \rho_X^\star \sigma_e \sigma_d \chi^{(X)} \sqrt{r_2}
\right)^2
\Bigg]\\
& - \frac{\eta^2 {(\nu^{(N)})}^2 {\rho_X^\star}^2 \sigma_c^2 \sigma_e^2 \sigma_f^2}{r_2}
\, \phi_N \phi_R =0,
\end{aligned}
\label{eq:instab}
\end{equation}
where $\rho_R^\star$ and $\rho_X^\star$ represent the critical resource and predator reciprocities at which instability sets in, respectively. Together with the cavity self-consistency equations, this condition defines a stability boundary curve in the phase space of reciprocity parameters $(\rho_R, \rho_X)$. When the system crosses this boundary, the unique self-averaging steady state becomes unstable, and the ecosystem transitions to a dynamic phase with chaotic fluctuations. Fig.~\ref{fig:stabboundary_si} shows this stability boundary as a solid black line, which agrees well with numerical simulations showing the probability of reaching a steady state as a heatmap.

\begin{figure}[ht!]
    \centering
    \includegraphics[width=0.6\textwidth]{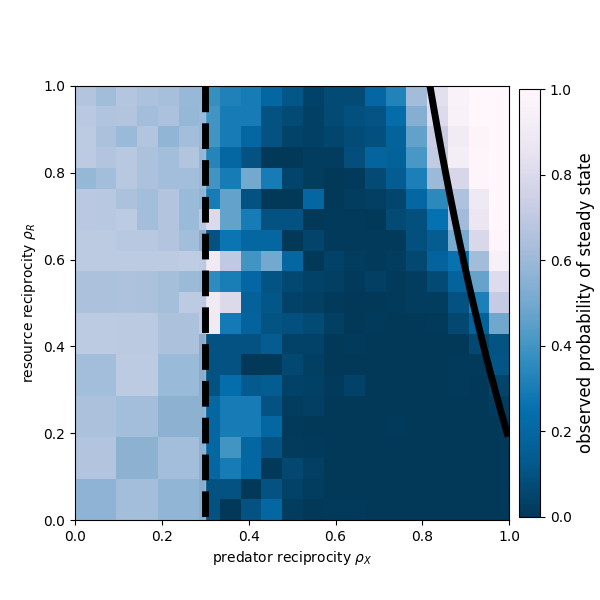} 
    \caption{\justifying Heatmap showing the fraction of simulations that reach a steady state. The solid curve denotes the instability boundary obtained from the cavity solution, while the dashed curve indicates the infeasibility boundary. The theoretical boundaries show good agreement with the numerical simulations.}
    \label{fig:stabboundary_si}
    
\end{figure}

This general instability condition is rather tedious, but it can be simplified for the case when $\sigma_c = \sigma_d = \sigma_e = \sigma_f = \sigma$, and equal species pool ratios
$r_1 = r_2 = 1$. In this case, solving for the  self-susceptibilities
$\nu^{(N)}$, $\chi^{(X)}$, and $K^{(R)}$, using the self consistency-equations~\eqref{eq:selfcon1}--\eqref{eq:selfcon2}, we obtain:
\begin{equation}
\nu^{(N)}
=
- \frac{\phi_N}{
\displaystyle
\frac{\eta \sigma^{2} \rho_R \phi_R}{1 - \eta \sigma^{2} \rho_R \nu^{(N)}}
- \frac{\phi_X}{\nu^{(N)}}
}.
\end{equation}

Solving this for $\nu^{(N)}$ we get:
\begin{equation}
\nu^{(N)}
=
\frac{\phi_X - \phi_N}{
\eta \sigma^{2} \rho_R \left( \phi_R + \phi_X - \phi_N \right)
}.
\end{equation}
And similarly we get $\chi^{(X)}$ and $K^{(R)}$ in terms of diversity and reciprocity.
\begin{equation}
\chi^{(X)}
=
\frac{\rho_R\left( \phi_R + \phi_X - \phi_N \right)\phi_X}{\rho_X(\phi_X - \phi_N)}
\end{equation}
\begin{equation}
K^{(R)}=    \left( \phi_R + \phi_X - \phi_N \right)
\end{equation}
By substituting these self-susceptibilities into the full instability condition Eq.~\eqref{eq:instab}, along with our simplifications of equal $\sigma$'s and $r_1=r_2=1$, we obtain a much simpler analytic expression for the instability boundary, which is Eq.~(6) in the main text, as follows:
\begin{equation}
{\rho_R^\star}^2 \phi_R^2 \phi_N (\phi_N\rho_X^2 - \phi_X) = \phi_N \phi_R \rho_X^2{(\phi_X-\phi_N)}^2 
\label{eq:stabcond}
\end{equation}\\
We can now use this equation to get the critical reciprocities $\rho^\dagger$. At $\rho_R^\star=1$, we get:
\begin{equation}
    \rho_X^\dagger = \sqrt{\frac{\phi_R \phi_X}{\phi_R \phi_N - (\phi_X - \phi_N)^2}}
    \label{eq:critX}
\end{equation}\\

Similarly, at $\rho_X^\star=1$, we get:
\begin{equation}
    \rho_R^\dagger = \sqrt{\frac{\phi_N - \phi_X}{\phi_R}}
    \label{eq:critR}
\end{equation}

The critical reciprocities $\rho_X^\dagger$ and $\rho_R^\dagger$, represent the corners of the instability boundary, and as we see they are controlled entirely by the diversity $\phi$ at each level. \\

\section*{Appendix F: Relationship between instability boundary and energy flow efficiency}

The energy conversion efficiency $\eta$ determines how effectively biomass is transferred between adjacent trophic levels. In the main text, we show that $\eta$ acts as a control parameter that rotates and translates the stability boundary, transitioning ecosystems between top-down and bottom-up control of stability. In this appendix, we analyze how $\eta$ modulates the stability boundary and the relative sensitivity of the ecosystem to nonreciprocal interactions at different trophic interfaces.


The critical reciprocities $\rho^\dagger_X$ and $\rho^\dagger_R$, derived in Appendix~E, depend implicitly on $\eta$ through the steady-state diversities $\phi_R$, $\phi_N$, and $\phi_X$. From Eqs.~\eqref{eq:critX} and \eqref{eq:critR}, we see that the critical reciprocities are controlled entirely by the fractions of surviving species at each trophic level. Since the diversities themselves depend on $\eta$ through the self-consistency equations~\eqref{eq:fullselfcon1}--\eqref{eq:selfcon2}, the stability boundary inherits a nontrivial dependence on energy flow efficiency.


Decreasing $\eta$ intensifies competitive exclusion at each trophic step, as less energy is available to support higher trophic levels. This steepens the diversity gradient $\phi_R > \phi_N > \phi_X$ that is already present when $\eta = 1$ (Fig.~\ref{fig:diversitieswitheta}). The steepened gradient has a pronounced effect on the stability boundary.

As shown in Fig.~\ref{fig:critreceta}, decreasing $\eta$ causes $\rho^\dagger_X$ to decrease monotonically toward zero, while $\rho^\dagger_R$ exhibits nonmonotonic behavior, first increasing and then decreasing. The ratio $\rho^\dagger_X/\rho^\dagger_R$ therefore decreases with $\eta$, indicating that at low efficiency, the ecosystem becomes more sensitive to nonreciprocal interactions at the resource--consumer interface than at the predator--consumer interface. In ecological terminology, this represents a transition from top-down to bottom-up control of stability. The intersection of the two curves in Fig.~\ref{fig:critreceta} marks the point at which the stability boundary becomes symmetric with respect to the two reciprocity parameters. For $\eta$ values above this intersection, $\rho^\dagger_X > \rho^\dagger_R$, indicating that stability is more sensitive to predator-level nonreciprocity (top-down control). Below the intersection, the inequality reverses, and resource-level nonreciprocity becomes the dominant driver of instability (bottom-up control).


A key consequence of decreasing $\eta$ is the expansion of the stable region in $(\rho_X, \rho_R)$ parameter space (Fig.~\ref{fig:boundarieswitheta}). This can be understood as follows. When $\eta$ is small, predator abundances are suppressed due to inefficient energy transfer, leading to $\phi_X \ll \phi_N$. From Eq.~\eqref{eq:critR}, this implies $\rho^\dagger_R$ approaches $\sqrt{\phi_N/\phi_R}$, which increases as the consumer-to-resource diversity ratio grows. Simultaneously, from Eq.~\eqref{eq:critX}, $\rho^\dagger_X$ decreases as $\phi_X$ diminishes relative to $\phi_N$. The net effect is that the stability boundary shifts to encompass a larger region of parameter space, indicating that ecosystems with low energy transfer efficiency are more robust to nonreciprocal interactions. This counterintuitive result arises because low efficiency naturally suppresses the top trophic level, reducing the strength of top-down feedback loops that can destabilize the system.

Fig.~\ref{fig:boundarieswitheta} displays the stability boundary for several values of $\eta$ ranging from 0.1 to 1.0. As $\eta$ decreases, the stable region (below and to the right of each curve) expands progressively. 
Fig.~\ref{fig:diversitieswitheta} shows the diversities or survival fractions $\phi_X$, $\phi_N$, and $\phi_R$ at the critical reciprocity values $\rho^\dagger_X$ and $\rho^\dagger_R$ as functions of $\eta$. Even at perfect efficiency, $\eta = 1$, there is a gradient in diversity, with $\phi_R\sim1$, $\phi_N$ lower and $\phi_X$ lower still. As $\eta$ decreases, the diversity gradient steepens: $\phi_R$ remains close to 1, $\phi_N$ decreases moderately, and $\phi_X$ drops precipitously. This steepening gradient explains both the expansion of the stable region and the shift from top-down to bottom-up control.

\begin{figure}[ht!]
    \centering
    \includegraphics[width=0.6\linewidth]{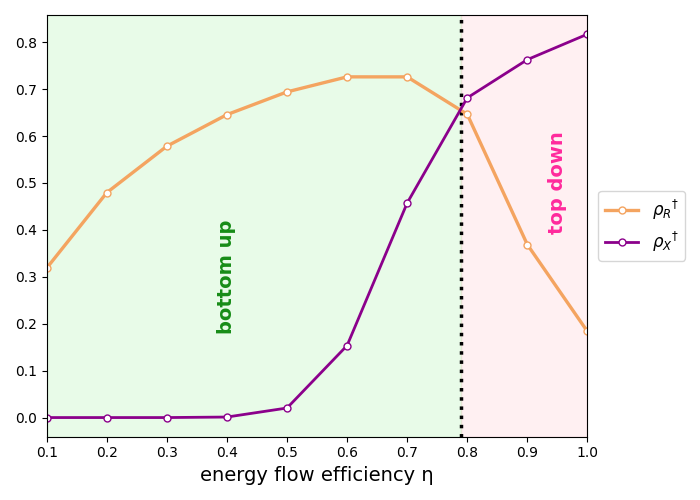}
    \caption{\justifying With decreasing $\eta$, $\rho_X^{\dagger}$ decreases toward zero, whereas
$\rho_R^{\dagger}$ exhibits a nonmonotonic dependence, first increasing and
then decreasing. This behavior leads to an expansion of the stability region
as $\eta$ decreases. The intersection of the two curves marks the point at
which the stability boundary is symmetric.
 }
 \label{fig:critreceta}
\end{figure}
\begin{figure}
    \centering
    \includegraphics[width=0.55\linewidth]{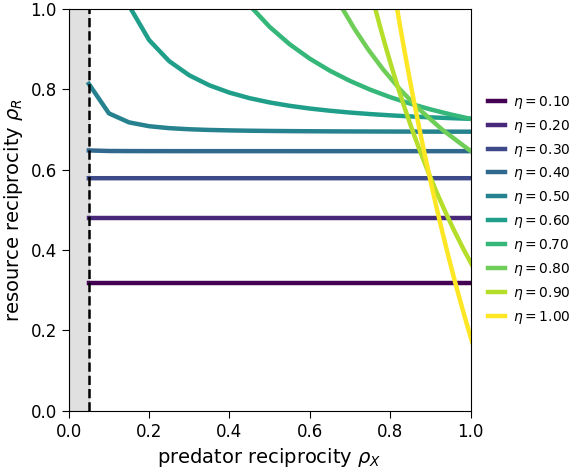}
    \caption{\justifying The stability boundary is shown for different values of the energy flow efficiency 
$\eta$. As $\eta$ decreases, the area of the stable region increases, indicating 
enhanced dynamical stability at lower energy transfer efficiencies. At the same time, 
the infeasible region contracts with decreasing $\eta$, reflecting a progressive 
relaxation of feasibility constraints. The grey shaded region at $\rho_X = 0$ denotes 
a universally infeasible regime: in the limit $\rho_x \to 0$, the susceptibility 
$\nu^{(N)}$ diverges, rendering the solution nonphysical for all values of $\eta$.
}
\label{fig:boundarieswitheta}
\end{figure}
\begin{figure}[ht!]
    \centering
    \includegraphics[width=0.65\linewidth]{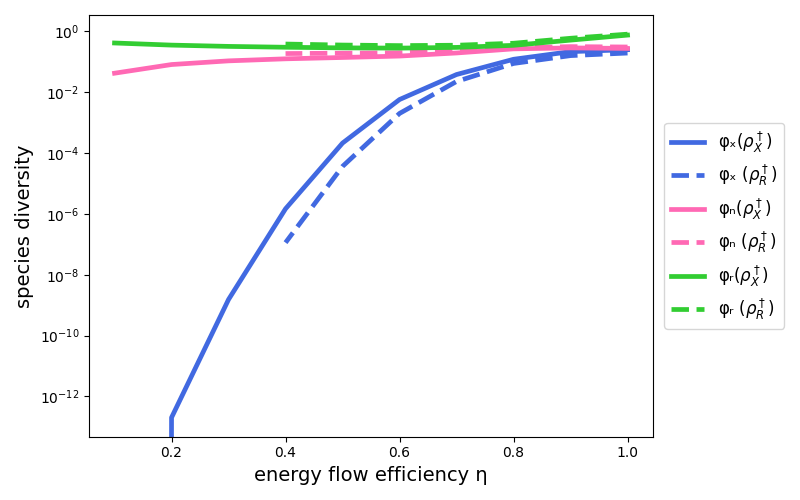}
    \caption{\justifying The diversities $\phi_X$, $\phi_N$, and $\phi_R$, computed at the critical reciprocity 
values $\rho_X^{\dagger}$ and $\rho_R^{\dagger}$, are shown as functions of the energy 
flow efficiency $\eta$. As we can see decreasing $\eta$ steepens diversity gradient $\phi_R > \phi_N > \phi_X$ .}
\label{fig:diversitieswitheta}
\end{figure}

\newpage

\section*{Appendix G: Relationship between stability and species pool sizes}

In the main text and preceding appendices, we assumed equal species pool sizes at each trophic level: $M_R = M_N = M_X$. Here, we relax this assumption and analyze how the ratios of pool sizes across trophic levels affect the stability boundary. We find that pyramidal pool structures---where lower trophic levels contain more species than higher levels---substantially expand the stable region. We parameterize the relative sizes of species pools using two ratios:
\begin{equation}
r_1 = \frac{M_N}{M_R}, \qquad r_2 = \frac{M_X}{M_N},
\end{equation}
where $M_R$, $M_N$, and $M_X$ are the number of resource, consumer, and predator species in the regional pool, respectively. The point $(r_1, r_2) = (1, 1)$ corresponds to equal pool sizes across all levels, which we analyzed in the main text. Varying these ratios allows us to explore different trophic structures. (Fig.~\ref{fig:pyramid}) schematically shows that  the $(r_1, r_2)$ plane can be divided into four qualitatively distinct regimes based on the ordering of pool sizes: 

\begin{enumerate}
    \item $r_1 < 1$, $r_2 < 1$: Pyramidal structure ($M_R > M_N > M_X$);
    \item $r_1 > 1$, $r_2 > 1$: Inverted pyramid ($M_R < M_N < M_X$);
    \item $r_1 < 1$, $r_2 > 1$: Hourglass structure ($M_N < M_R$ and $M_N < M_X$); and
    \item $r_1 > 1$, $r_2 < 1$: Diamond structure ($M_N > M_R$ and $M_N > M_X$).
\end{enumerate}

\begin{figure}
    \centering
    \includegraphics[width=0.6\linewidth]{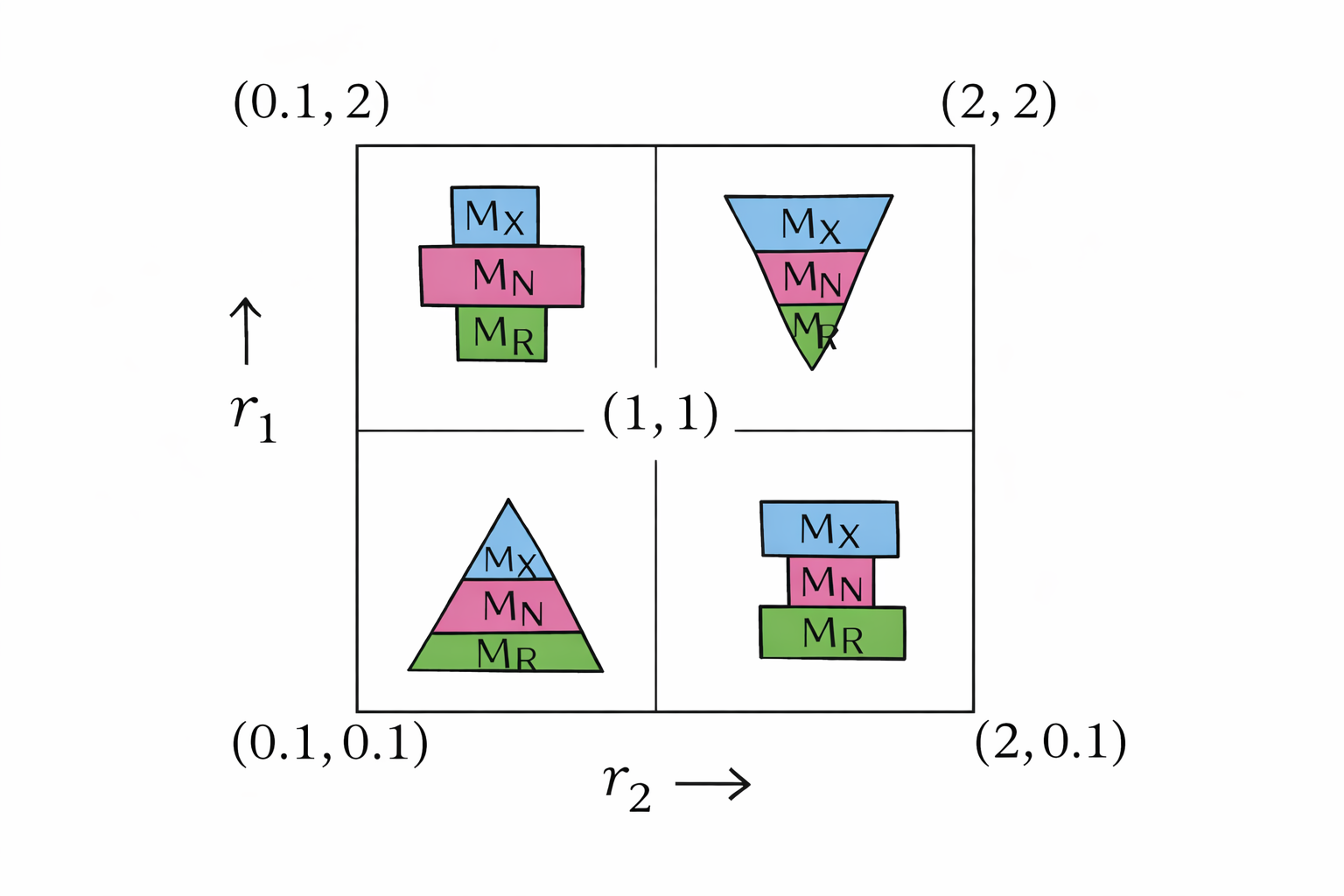}
    \caption{\justifying 
Schematic map of pool-size ratio regimes in the $(r_1,r_2)$ plane, where
$r_1 = M_N/M_R$ and $r_2 = M_X/M_N$.
Each quadrant corresponds to a distinct ordering of the relative pool sizes
$M_R$, $M_N$, and $M_X$.
The point $(1,1)$ denotes equal pool sizes.}
    \label{fig:pyramid}
\end{figure}
    
Fig.~\ref{fig:r1r2area} shows how the stability boundary in the $(\rho_X, \rho_R)$ plane changes as we vary the pool size ratios $r_1$ and $r_2$ together. Several key patterns emerge.

First, decreasing both $r_1$ and $r_2$ (moving toward pyramidal structure) causes the stability boundary to shift, enlarging the stable region. This can be understood through the cavity equations: smaller $r_1$ and $r_2$ reduce the effective pressure from higher trophic levels on the consumer layer, which appears in the susceptibility equations (Eqs.~74--76). With fewer predators relative to consumers and fewer consumers relative to resources, the feedback loops that drive instability are weakened.

Second, the asymmetry between $\rho_X$ and $\rho_R$ sensitivity is modulated by the pool size ratios. When $r_2 \ll 1$ (few predators relative to consumers), the critical reciprocity $\rho^\dagger_X$ decreases, indicating reduced sensitivity to predator-level nonreciprocity. Conversely, when $r_1 \ll 1$ (few consumers relative to resources), the system becomes less sensitive to resource-level nonreciprocity. The interplay between $r_1$ and $r_2$ thus provides additional control over whether stability is top-down or bottom-up controlled.

Third, increasing both ratios toward $r_1 > 1$ and $r_2 > 1$ (inverted pyramid) progressively shrinks the stable region. In the limit of large $r_1$ and $r_2$, the stable region contracts significantly, indicating that top-heavy trophic structures are intrinsically prone to instability.

\begin{figure}[ht!]
    \centering
    \includegraphics[width=0.65\linewidth]{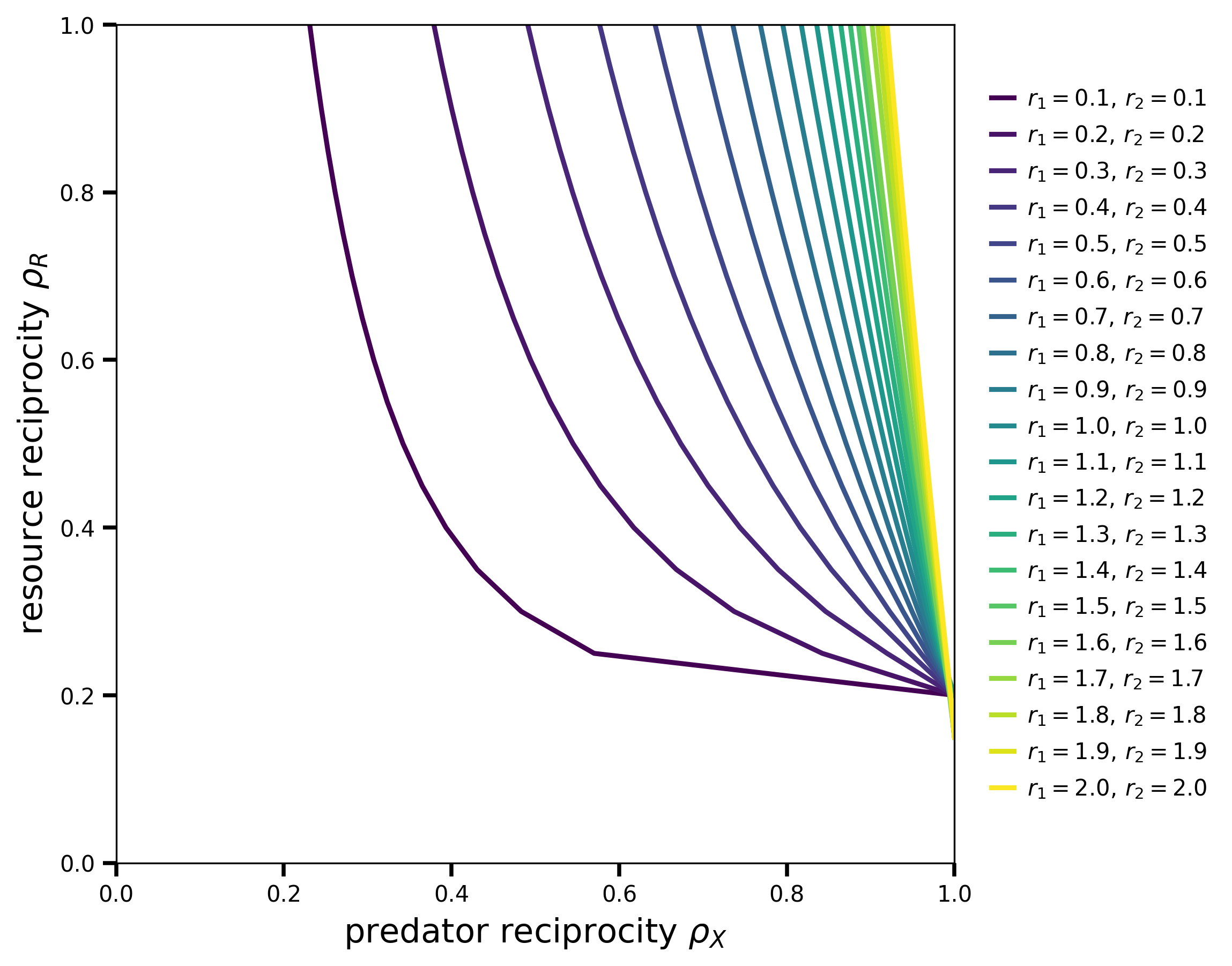}
    \caption{\justifying Stability boundaries in the \((\rho_X,\rho_R)\) plane for different pool-size ratios
\(r_1 = M_N/M_R\) and \(r_2 = M_X/M_N\).
Decreasing \(r_1\) and \(r_2\) enlarges the stable region and suppresses top--down control.
}
\label{fig:r1r2area}
\end{figure}

To quantify the overall effect of trophic structure on stability, we computed the total area of the stable region in the $(\rho_X, \rho_R)$ plane across a range of $(r_1, r_2)$ values (Fig.~\ref{fig:arear1r2}). The stable area exhibits a pronounced gradient: it is largest in the lower-left corner of the $(r_1, r_2)$ plane (pyramidal structure) and smallest in the upper-right corner (inverted pyramid).

This result has a natural interpretation. In pyramidal ecosystems, the large resource base provides a stable foundation that buffers against fluctuations propagating up the food web. The progressively smaller pools at higher levels mean that predator dynamics, while potentially volatile, affect fewer species and have diminished impact on overall ecosystem stability. In contrast, inverted pyramids concentrate species at higher trophic levels, amplifying the destabilizing effects of top-down feedback loops.

The heatmap in Fig.~\ref{fig:arear1r2} reveals that the stable area varies by more than an order of magnitude across the range of pool size ratios considered. This substantial variation underscores the importance of trophic structure---not just interaction strengths or reciprocity---in determining ecosystem stability.




    
\begin{figure}[ht!]
    \centering
    \includegraphics[width=0.75\linewidth]{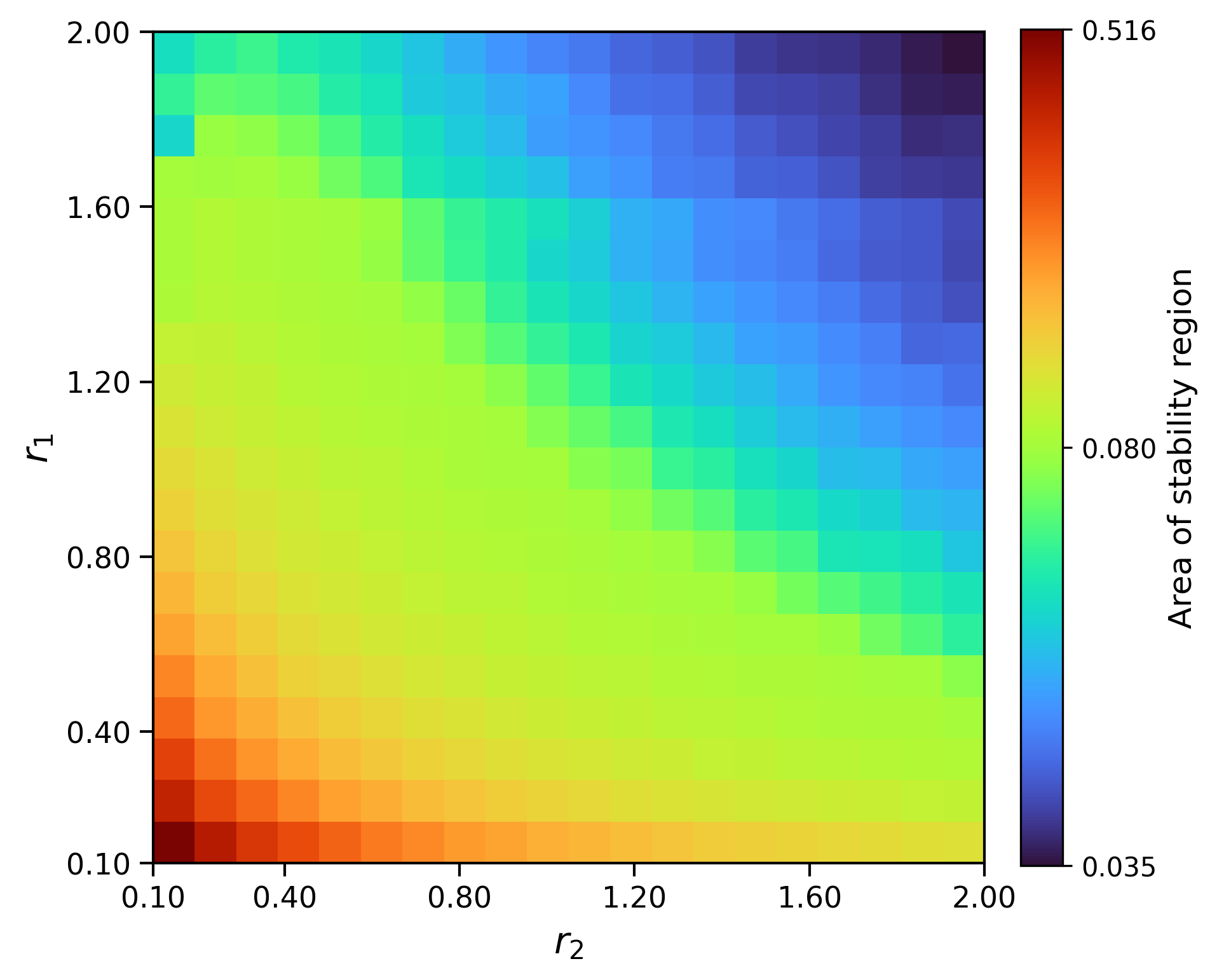}
    \caption{\justifying Heatmap showing the area of the stability region as a function of the pool-size ratios
\( r_1 = M_N/M_R \) and \( r_2 = M_X/M_N \).
The central color corresponds to the symmetric case \( r_1 = r_2 = 1 \), which
recovers the original model.
As both \( r_1 \) and \( r_2 \) decrease, the stability region expands.
}
\label{fig:arear1r2}
\end{figure}

\section*{Appendix H: Generalization to $N$ trophic levels}
\noindent
In the main text, we analyzed ecosystems with three trophic levels. Here, we generalize the model and stability analysis to ecosystems with $N$ trophic levels, demonstrating that the key results---the dependence of stability on diversity gradients across levels---extend to arbitrarily complex food webs.

We consider an ecosystem with $N$ trophic levels, indexed by $i = 1, 2, \ldots, N$. Level $i = 1$ corresponds to basal resources (primary producers), while level $i = N$ corresponds to apex predators. Each level $i$ contains $M_i$ species, and we denote the abundance of species $\mu$ at level $i$ by $B_\mu^{(i)}$.

\subsubsection{Dynamics}

The dynamical equations generalize the three-level model presented in the main text. Species at the top level ($i = N$) gain energy only from the level below and experience mortality:
\begin{equation}
\frac{1}{B_\mu^{(N)}}\frac{d B_\mu^{(N)}}{dt}
=
\eta \sum_{\nu} \alpha_{\nu\mu}^{(N,N-1)} \, B_\nu^{(N-1)}
- m_\mu^{(N)}.
\end{equation}

Species at intermediate levels ($i = 2, \ldots, N-1$) gain energy from the level below, lose biomass to predation from the level above, and experience mortality:
\begin{equation}
\frac{1}{B_\mu^{(i)}}\frac{d B_\mu^{(i)}}{dt}
=
\eta \sum_{\nu} \alpha_{\nu\mu}^{(i,i-1)} \, B_\nu^{(i-1)}
-
\sum_{\nu} \beta_{\nu\mu}^{(i+1,i)} \, B_\nu^{(i+1)}
-
m_\mu^{(i)}.
\end{equation}

Species at the basal level ($i = 1$) follow logistic growth in the absence of consumers, with self-regulation encoded in the $-B_\mu^{(1)}$ term:
\begin{equation}
\frac{1}{B_\mu^{(1)}}\frac{d B_\mu^{(1)}}{dt}
=
K_\mu^{(1)}
-
B_\mu^{(1)}
-
\sum_{\nu} \beta_{\nu\mu}^{(2,1)} \, B_\nu^{(2)}.
\end{equation}

\subsubsection{Parametrization and ensemble}

Following the parameterization used in the main text, we specify the interaction matrices to ensure a proper thermodynamic limit as $M_i \to \infty$. The consumption coefficients, encoding the energy flux from level $i-1$ to level $i$, are given by:
\begin{equation}
\alpha_{\mu\nu}^{(i,i-1)}
=
\frac{\mu_\alpha^{(i)}}{M_i}
+
\frac{\sigma_\alpha^{(i)}}{\sqrt{M_i}} \, d_{\mu\nu}^{(i)},
\end{equation}
where the random variables $d_{\mu\nu}^{(i)}$ satisfy:
\begin{equation}
\langle d_{\mu\nu}^{(i)} \rangle = 0,
\qquad
\langle d_{\mu\nu}^{(i)} \, d_{\alpha\beta}^{(i)} \rangle
=
\delta_{\mu\alpha}\,\delta_{\nu\beta}.
\end{equation}

The predation coefficients, encoding the depletion of level $i$ by level $i+1$, are parameterized with a correlation $\rho_{i+1,i}$ to the consumption coefficients:
\begin{equation}
\beta_{\nu\mu}^{(i+1,i)}
=
r_{i+1,i}^{-1}
\left(
\frac{\mu_\beta^{(i)}}{M_i}
+
\frac{\sigma_\beta^{(i)} }{\sqrt{r_{i+1,i}}\sqrt{M_i}}
\left(
\rho_{i+1,i} \, d_{\nu\mu}^{(i)}
+
\sqrt{1-\rho_{i+1,i}^2}\, g_{\nu\mu}^{(i)}
\right)
\right),
\end{equation}
where the additional random variables $g_{\nu\mu}^{(i)}$ are independent of $d_{\nu\mu}^{(i)}$ and satisfy:
\begin{equation}
\langle g_{\nu\mu}^{(i)} \rangle = 0,
\qquad
\langle g_{\nu\mu}^{(i)} \, g_{\alpha\beta}^{(i)} \rangle
=
\delta_{\nu\alpha}\,\delta_{\mu\beta}.
\end{equation}

The correlation parameter $\rho_{i+1,i}$ controls the degree of reciprocity at the interface between levels $i$ and $i+1$. This generalizes the stratified nonreciprocity of the three-level model: with $N$ levels, there are $N-1$ independent reciprocity parameters, one at each trophic interface.


We define the ratios of pool sizes between adjacent trophic levels:
\begin{equation}
r_{N,N-1} = \frac{M_N}{M_{N-1}}, \qquad
\ldots \qquad
r_{2,1} = \frac{M_2}{M_1}.
\end{equation}

The mortality rates at each level are drawn from distributions with level-specific means and variances:
\begin{equation}
m_\mu^{(i)}
=
m_i
+
\delta m_\mu^{(i)} ,
\qquad
\langle \delta m_\mu^{(i)} \rangle = 0,
\qquad
\langle \delta m_\mu^{(i)} \, \delta m_\nu^{(i)} \rangle
=
(\sigma_m^{(i)})^2 \, \delta_{\mu\nu}.
\end{equation}

The carrying capacities at the basal level are similarly drawn:
\begin{equation}
K_\mu^{(1)}
=
K_1
+
\delta K_\mu^{(1)},
\qquad
\langle \delta K_\mu^{(1)} \rangle = 0,
\qquad
\langle \delta K_\mu^{(1)} \, \delta K_\nu^{(1)} \rangle
=
\delta_{\mu\nu}.
\end{equation}

\subsubsection{Cavity solution}

Applying the cavity method to the $N$-level model, we can show that species abundances at each trophic level generally follow truncated Gaussian distributions. The structure of these distributions reflects the position of each level within the food web.

At the top level ($i = N$), species receive energy only from below:
\begin{equation}
B_0^{(N)}
=
\max\!\left\{
0,\,
\frac{
\eta \mu_\alpha^{(N)} \langle B^{(N-1)} \rangle
-
m_{N}
+
\sqrt{
(\sigma_m^{(N)})^2
+
\eta^2 (\sigma_\alpha^{(N)})^2 \langle (B^{(N-1)})^2 \rangle
}
\, Z_N
}{-
\eta \sigma_\alpha^{(N)} \sigma_\beta^{(N-1)}
\rho_{N,N-1}
\,
S^{(N-1)}
\sqrt{r_{N,N-1}^{-1}}
}
\right\}.
\end{equation}

At intermediate levels ($i = 2, \ldots, N-1$), species experience both bottom-up energy gain and top-down predation pressure:
\begin{equation}
\begin{aligned}
B_0^{(i)}
&=
\max\!\Bigg\{
0,\,
\frac{
\eta \mu_\alpha^{(i)} \langle B^{(i-1)} \rangle
-
m_i
-
\mu_\beta^{(i)} \langle B^{(i+1)} \rangle
+
\sqrt{
(\sigma_m^{(i)})^2
+
\eta^2 (\sigma_\alpha^{(i)})^2 \langle (B^{(i-1)})^2 \rangle
+
(\sigma_\beta^{(i)})^2 \langle (B^{(i+1)})^2 \rangle
}
\, Z_i
}{
\eta \sigma_\alpha^{(i)} \sigma_\beta^{(i-1)}
\sqrt{r_{i,i-1}^{-1}}
\rho_{i,i-1}
S^{(i-1)}
-
\eta \sigma_\alpha^{(i+1)} \sigma_\beta^{(i)}
\sqrt{r_{i+1,i}}
\rho_{i+1,i}
S^{(i+1)}
}
\Bigg\}.
\end{aligned}
\end{equation}

At the basal level ($i = 1$), self-regulation provides an intrinsic carrying capacity:
\begin{equation}
B_0^{(1)}
=
\max\!\left\{
0,\,
\frac{
K^{(1)}
-
\mu_\beta^{(1)} \langle B^{(2)} \rangle
+
\sqrt{
(\sigma_K^{(1)})^2
+
(\sigma_\beta^{(1)})^2 \langle (B^{(2)})^2 \rangle
}
\, Z_1
}{
1
-
\eta \sigma_\alpha^{(1)} \sigma_\beta^{(1)}
\sqrt{r_{2,1}}
\rho_{2,1}
S^{(2)}
}
\right\}.
\end{equation}


The self-susceptibility $S^{(i)}$ characterizes how the abundance of species at level $i$ responds to perturbations in their own environmental parameters. These susceptibilities satisfy a system of coupled equations that propagate information up and down the food web.

At the top level ($i = N$):
\begin{equation}
S^{(N)}
=
\frac{
\phi_N
}{
\eta \, \sigma_\alpha^{(N)} \sigma_\beta^{(N-1)}
\, \rho_{N,N-1}
\sqrt{r_{N,N-1}^{-1}}
\,
S^{(N-1)}
}.
\end{equation}

At intermediate levels ($i = 2, \ldots, N-1$):
\begin{equation}
S^{(i)}
=
\frac{
-\phi_i
}{
\eta \, \sigma_\alpha^{(i)} \sigma_\beta^{(i-1)}
\sqrt{r_{i,i-1}^{-1}}
\, \rho_{i,i-1}
\, S^{(i-1)}
-
\eta \, \sigma_\alpha^{(i+1)} \sigma_\beta^{(i)}
\sqrt{r_{i+1,i}}
\, \rho_{i+1,i}
\, S^{(i+1)}
}.
\end{equation}

At the basal level ($i = 1$):
\begin{equation}
S^{(1)}
=
\frac{
\phi_1
}{
1
-
\eta \, \sigma_\alpha^{(1)} \sigma_\beta^{(2)}
\sqrt{r_{2,1}}
\, \rho_{2,1}
\, S^{(2)}
}.
\end{equation}

\subsubsection{Stability analysis}

Following the approach of Appendix~E, we analyze stability by computing the second moments of the susceptibilities under random perturbations. The loss of stability is signaled by the divergence of these moments. Perturbing each level and computing the response, we obtain:

For the top level ($i = N$):
\begin{equation}
\left\langle \left( \frac{d B_0^{(N)+}}{d\varepsilon} \right)^2 \right\rangle
=
\frac{
\phi_{N-1}\,(\sigma_\alpha^{(N)})^2
}{
(\sigma_\alpha^{(N)})^2
(\sigma_\beta^{(N-1)})^2
(\rho_{N,N-1})^2
(S^{(N-1)})^2
r_{N,N-1}^{-1}
}
\left[
\left\langle
\left( \frac{d B_0^{(N-1)+}}{d\varepsilon} \right)^2
\right\rangle
+ 1
\right].
\end{equation}

For intermediate levels ($i = 2, \ldots, N-1$):
\begin{align}
\left\langle \left( \frac{d B_0^{(i)+}}{d\varepsilon} \right)^2 \right\rangle
&=
\frac{1}{
\Big(
\sigma_\alpha^{(i)}\sigma_\beta^{(i-1)}
S^{(i-1)}\rho_{i,i-1}\sqrt{r_{i,i-1}^{-1}}
-
\rho_{i+1,i}\sigma_\beta^{(i)}\sigma_\alpha^{(i+1)}
S^{(i+1)}\sqrt{r_{i+1,i}}
\Big)^2
}
\nonumber\\
&\quad\times
\Bigg\{
(\sigma_\alpha^{(i)})^2 \phi_{i-1}
\left[
\left\langle
\left( \frac{d B_0^{(i-1)+}}{d\varepsilon} \right)^2
\right\rangle
+ 1
\right]
\nonumber\\
&\qquad\quad
+
(\sigma_\beta^{(i)})^2 \phi_{i+1}
\left[
\left\langle
\left( \frac{d B_0^{(i+1)+}}{d\varepsilon} \right)^2
\right\rangle
+ 1
\right]
\Bigg\}.
\end{align}

For the basal level ($i = 1$):
\begin{align}
\left\langle \left( \frac{d B_0^{(1)+}}{d\epsilon} \right)^2 \right\rangle
&=
\frac{
(\sigma_\beta^{(1)})^2
\phi_2
\left(
\left\langle
\left( \frac{d B_0^{(2)+}}{d\epsilon} \right)^2
\right\rangle
+ 1
\right)
}{
\left(
1
-
\sqrt{r_{21}}
\sigma_\beta^{(1)}
\sigma_\alpha^{(2)}
\rho_{21}
S^{(2)}
\right)^2
}.
\end{align}

\subsubsection{Instability condition}

The coupled equations for the susceptibility second moments can be written in matrix form as $\mathbf{A}\mathbf{X} = \mathbf{B}$, where $\mathbf{X}$ is the vector of susceptibility second moments:
\begin{equation}
\mathbf{X}=
\begin{pmatrix}
\left\langle\left(\frac{d B_0^{(1)+}}{d\varepsilon}\right)^2\right\rangle  \\[6pt]
\vdots \\[6pt]
\left\langle\left(\frac{d B_0^{(i)+}}{d\varepsilon}\right)^2\right\rangle  \\[6pt]
\vdots \\[6pt]
\left\langle\left(\frac{d B_0^{(N)+}}{d\varepsilon}\right)^2\right\rangle 
\end{pmatrix},
\end{equation}
and $\mathbf{B}$ is the inhomogeneous term:
\begin{equation}
\mathbf{B}= \begin{pmatrix}
(\sigma_\beta^{(1)})^2 \phi_2 \\[6pt]
\vdots \\[6pt]
\eta (\sigma_\alpha^{(i)})^2 \phi_{i-1}+(\sigma_\beta^{(i)})^2 \phi_{i+1} \\[6pt]
\vdots \\[6pt]
\eta \phi_{N-1}(\sigma_\alpha^{(N)})^2
\end{pmatrix}.
\end{equation}

The matrix $\mathbf{A}$ is tridiagonal, reflecting the fact that interactions are limited to adjacent trophic levels:
\begin{equation}
A_{ij}=0 \qquad \text{for } |i-j|>1.
\end{equation}

For interior levels ($i \neq 1, N$), the matrix elements are:
\begin{align}
A_{ii}
&=
\Big(
\sigma_{\alpha}^{(i)} \sigma_{\beta}^{(i-1)} S^{(i-1)}
\rho_{i,i-1}\sqrt{r_{i,i-1}^{-1}}
-
\rho_{i+1,i}\,
\sigma_{\beta}^{(i)} \sigma_{\alpha}^{(i+1)} S^{(i+1)}
\sqrt{r_{i+1,i}}
\Big)^2,
\\[6pt]
A_{i,i+1}
&=
- \big(\sigma_{\beta}^{(i)}\big)^2 \, \phi_{i+1},
\\[6pt]
A_{i,i-1}
&=
- \eta \, \big(\sigma_{\alpha}^{(i)}\big)^2 \, \phi_{i-1}.
\end{align}

For the boundary levels, the elements are:
\begin{align}
A_{11}
&=
\left(
1
-
\sqrt{r_{21}}\,
\sigma_{\beta}^{(1)} \sigma_{\alpha}^{(2)}
\rho_{21}\, S^{(2)}
\right)^2,
\\[6pt]
A_{12}
&=
- \big(\sigma_{\beta}^{(1)}\big)^2 \, \phi_2,
\\[6pt]
A_{NN}
&=
\big(\sigma_{\alpha}^{(N)}\big)^2
\big(\sigma_{\beta}^{(N-1)}\big)^2
\big(\rho_{N,N-1}\big)^2
\big(S^{(N-1)}\big)^2
\, r_{N,N-1}^{-1},
\\[6pt]
A_{N,N-1}
&=
- \eta \, \phi_{N-1} \big(\sigma_{\alpha}^{(N)}\big)^2.
\end{align}


The susceptibility second moments diverge when $\det(\mathbf{A}) = 0$. For a tridiagonal matrix, the determinant can be computed recursively. Defining $D_n \equiv \det(\mathbf{A}_n)$, where $\mathbf{A}_n$ is the $n \times n$ principal submatrix, we have:
\begin{align}
D_0 &= 1, \\[4pt]
D_1 &= A_{11}, \\[4pt]
D_n &= A_{nn}\, D_{n-1}
      - A_{n-1,n}\, A_{n,n-1}\, D_{n-2},
      \qquad n \ge 2.
\end{align}

\subsubsection{Recovery of three-level result}

To verify consistency with the main text, we specialize to the case $N = 3$ with all interaction strengths equal ($\sigma_\alpha^{(i)} = \sigma_\beta^{(i)} = \sigma$ for all $i$). The matrix $\mathbf{A}$ becomes:
\begin{equation}
\mathbf{A} =
\begin{pmatrix}
\left(
1 - \eta \sqrt{r_{21}}\,\sigma^{2}\rho_{21}\,S^{(2)}
\right)^2
& - \sigma^{2}\phi_{2}
& 0
\\[8pt]
- \eta^2 \,\phi_{1}\sigma^{2}
&
\eta^2 \left(
\sigma^{2} S^{(1)}\rho_{21}\sqrt{r_{21}^{-1}}
-
\rho_{32}\sigma^{2} S^{(3)}\sqrt{r_{32}}
\right)^2
& - \sigma^{2}\phi_{3}
\\[8pt]
0
& - \eta^2 \,\phi_{2}\sigma^{2}
&
\eta^2 \sigma^{4}\rho_{32}^{\,2}\big(S^{(2)}\big)^{2} r_{32}^{-1}
\end{pmatrix}.
\end{equation}

Applying the recursion relation:
\begin{align}
D_0 &= 1, \\[6pt]
D_1 &= 
\left(
1 - \sqrt{r_{21}}\,\sigma^{2}\rho_{21}\,S^{(2)}\,\eta
\right)^{2}, \\[6pt]
D_2 &= 
\left(\eta 
\sigma^{2} S^{(1)} \rho_{21}\sqrt{r_{21}^{-1}}
-
\eta \rho_{32}\sigma^{2} S^{(3)}\sqrt{r_{32}}
\right)^{2}
\left(
1 - \sqrt{r_{21}}\,\sigma^{2}\rho_{21}\,S^{(2)}\,\eta
\right)^{2}
- \sigma^{4}\,\phi_{1}\phi_{2}\,\eta^{2},
\end{align}
and
\begin{align}
D_3 &=
\eta^{2}\sigma^{4}\rho_{32}^{2}\big(S^{(2)}\big)^{2} r_{32}^{-1}
\Bigg[
\left(
\eta \sigma^{2} S^{(1)} \rho_{21}\sqrt{r_{21}^{-1}}
-
\eta \rho_{32}\sigma^{2} S^{(3)}\sqrt{r_{32}}
\right)^{2}
\left(
1 - \sqrt{r_{21}}\,\sigma^{2}\rho_{21}\,S^{(2)}\,\eta
\right)^{2}
\nonumber\\
&\qquad
- \sigma^{4}\phi_{1}\phi_{2}\eta^{2}
\Bigg]
- \sigma^{4}\phi_{3}\phi_{2}\eta^{2}
\left(
1 - \sqrt{r_{21}}\,\sigma^{2}\rho_{21}\,S^{(2)}\,\eta
\right)^{2}.
\end{align}

Setting $D_3 = 0$ yields the stability condition:
\begin{align}
&\left(
1 - \sqrt{r_{21}}\,\sigma^{2}\rho_{21}\,S^{(2)}\eta
\right)^{2}
\Bigg[
- \phi_{3}\phi_{2}
+ \frac{\big(S^{(2)}\big)^{2}\rho_{32}^{2}}{r_{32}}
\left(
\frac{\eta\sigma^{2} S^{(1)}\rho_{21}}{\sqrt{r_{21}}}
- \eta\rho_{32}\sigma^{2} S^{(3)}\sqrt{r_{32}}
\right)^{2}
\Bigg]
\nonumber\\
&\qquad\qquad
-
\frac{\eta^{2}\rho_{32}^{2}\big(S^{(2)}\big)^{2}}{r_{32}}
\,\phi_{2}\phi_{1}\,\sigma^{4} = 0.
\end{align}

Identifying variables with the notation of the main text:
\begin{equation}
S^{(1)} = K^{(R)}, 
\quad
S^{(2)} = \nu^{(N)}, 
\quad
S^{(3)} = \chi^{(X)},
\quad
\phi_{1} = \phi_{R},
\quad
\phi_{2} = \phi_{N},
\quad
\phi_{3} = \phi_{X},
\quad
\rho_{32} = \rho_{X},
\quad
\rho_{21} = \rho_{R},
\end{equation}
we recover the stability condition Eq.~\eqref{eq:stabcond} derived in Appendix~E for the three-level model. This confirms that the general $N$-level framework correctly reduces to our original results.

The tridiagonal structure of $\mathbf{A}$ reveals an important feature of multitrophic stability: perturbations propagate only between adjacent levels, but the stability condition couples all levels through the recursive determinant. This means that nonreciprocity at any interface can potentially destabilize the entire ecosystem, but its effect is mediated by the diversity gradient $\phi_1 > \phi_2 > \cdots > \phi_N$ established by directional energy flow.
\end{widetext}

\end{widetext}

\end{document}